\documentclass[aps,prl,twocolumn,superscriptaddress,nofootinbib,notitlepage,longbibliography]{revtex4-1}

\usepackage{amsmath,amssymb,amsfonts,bm,color,graphicx,tabularx}

\usepackage[etex=true,export]{adjustbox}
\usepackage{makecell}
\usepackage{multirow}
\usepackage{amsmath}
\usepackage{verbatim}
\usepackage{longtable}
\usepackage[colorlinks=true,linkcolor=blue,anchorcolor=red,citecolor=blue, urlcolor=blue]{hyperref}

\hypersetup{linkcolor=blue, citecolor=blue,urlcolor=blue}
\usepackage{tabularx,multirow,array,diagbox}
\usepackage{adjustbox}
\usepackage{supertabular}

\newcommand{\beq}{\begin{equation}}
\newcommand{\eeq}{\end{equation}}
\newcommand{\beql}{\begin{equation*}}
\newcommand{\eeql}{\end{equation*}}
\newcommand{\beqn}{\begin{eqnarray}}
\newcommand{\eeqn}{\end{eqnarray}}

\begin{document}
\title{ Spin Group Symmetry Criteria For Unconventional Magnetism }

\author{Xun-Jiang Luo}
\email{xjluo@hmfl.ac.cn}
\affiliation{Department of Physics, Hong Kong University of Science and Technology, Clear Water Bay, 999077 Hong Kong, China}
\affiliation{Anhui Province Key Laboratory of Low-Energy Quantum Materials and Devices, 
High Magnetic Field Laboratory, HFIPS, Chinese Academy of Sciences, Hefei, Anhui 
230031, China }

\author{Jin-Xin Hu}
\affiliation{Department of Physics, Hong Kong University of Science and Technology, Clear Water Bay, 999077 Hong Kong, China}

\author{Mengli Hu}
\affiliation{Institute for Theoretical Solid State Physics, IFW Dresden, 01069 Dresden, Germany}

\author{K. T. Law}
\email{phlaw@ust.hk}
\affiliation{Department of Physics, Hong Kong University of Science and Technology, Clear Water Bay, 999077 Hong Kong, China}

\begin{abstract}
Unconventional magnetism has typically been classified into two fundamental classes:  even-parity magnets (EPMs) and odd-parity magnets (OPMs). These two classes exhibit identical and opposite spin splittings, respectively, under momentum inversion, while both maintain symmetry-compensated magnetization. In this Letter, we present a unified spin space group-based framework that establishes comprehensive symmetry criteria for both classes. Our framework not only yields a complete classification of EPMs and OPMs but also uncovers a wealth of new symmetry-driven mechanisms for them. Specifically, we classify both classes into three types based on their spin textures: collinear (type-I), coplanar (type-II), and noncoplanar (type-III), and we demonstrate that both classes can be realized across collinear, coplanar, and noncoplanar magnetic orders. We identify eight distinct symmetry-driven mechanisms for OPMs and seven for EPMs, among which some paradigms of unconventional magnetism, for instance, altermagnets naturally emerge as one specific mechanism of EPMs.  Using these established criteria, we identify numerous candidate materials from the Magndata database, realizing some new symmetry mechanisms for OPMs and EPMs. Our work establishes a foundational symmetry framework for understanding, predicting, and designing unconventional magnetic materials.

\end{abstract}


\maketitle

\textit{Introduction.}---
Unconventional magnetism has emerged as a central theme in contemporary condensed matter physics \cite{Libor2020,Ma2021,Feng2022,Tomas2022,ifmmode2022,ifmmode2022a,Krempaský2024,ZhouXiaodong2024,Reimers2024,BaiLing2024,McClarty2024,LeeSuyoung2024,PhysRevLett.134.026001,Chen2025,Liu2025a,HuMengli2025,Zhu2025}. Such magnetic systems feature symmetry-compensated magnetization and non-relativistic spin splitting (NSS), which arises even in the absence of spin–orbit coupling (SOC). These characteristics make them highly promising for spintronics applications \cite{Jakub2021,ZhangRun-Wu2024,LeiHan2024,Zhou2025,Zhang2025,hu2025nonlinear}, as well as for exploring unconventional superconductivity \cite{PhysRevB.108.184505,Zhangsongbo2024,Ghorashi2024} and designing multifunctional materials \cite{DuanXunkai2025,GuMingqiang2025,96gy-sn83}. Unconventional magnetism has typically been classified into two fundamental classes:  even-parity magnets (EPMs) and odd-parity magnets (OPMs). These two magnetic classes exhibit identical and opposite spin splittings, respectively, under momentum inversion. The most well-known example of EPMs is altermagnetism, which features even-parity spin splitting (such as $d$- or $g$-wave) in collinear magnetic orders \cite{ifmmode2022a}. OPMs, in contrast, exhibit an odd-wave (such as $p$- or $f$-wave) splitting pattern that reverses sign under momentum inversion,  analogous to the effect of Rashba SOC but originating from purely non-relativistic mechanisms \cite{HayamiSatoru2020,BirkHellenes2023,Brekke2024,zk69-k6b2,85fd-dmy8,YuPing2025,2025Minghuan,Huang2025,2025li,2025zhu,zhuang2025,2025liu}.


The spin space group (SSG) offers a comprehensive symmetry framework for systems with negligible SOC \cite{LiuPengfei2022, CheXiaobing2024, JiangYi2024, XiaoZhenyu2024}. This framework has proven powerful in classifying unconventional magnetism \cite{Mazin2023,Sheng2024,LiuYichen2024,SunWei2025,2025song}, as well as in exploring external control of these materials \cite{DuanXunkai2025,GuMingqiang2025,96gy-sn83}. Despite rapid progress, a comprehensive symmetry-based understanding of OPMs and EPMs remains elusive, as their fundamental symmetry criteria have yet to be established within a unified framework. This gap has led to fragmented and potentially restrictive understandings: EPMs are predominantly associated with altermagnetism in collinear magnetic orders \cite{ifmmode2022}, while the realization of OPMs has traditionally been viewed as requiring \(T\tau\) symmetry (time reversal combined with a fractional translation) in coplanar magnetic orders \cite{BirkHellenes2023}. Such narrow perspectives have inadvertently constrained the exploration of unconventional magnetism, potentially overlooking a vast landscape of magnetic materials that do not conform to these specific paradigms. Consequently, establishing unified symmetry-based criteria for unconventional magnetism is an urgent necessity for the field. 


\begin{figure*}
\centering
\includegraphics[width=7in]{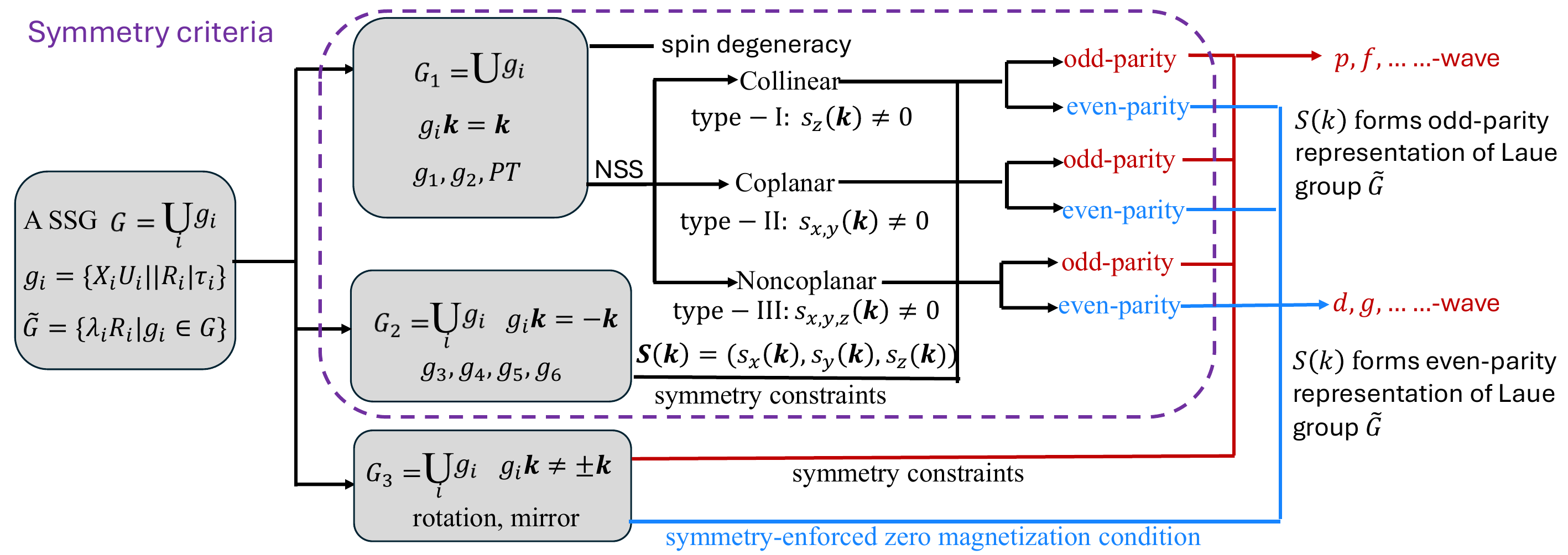}
\caption{Schematic illustration of the unified SSG framework for unconventional magnetism. From a SSG \(G\), symmetries preserving momentum (\(g_i\bm{k}=\bm{k}\)) constrain the spin texture dimensionality, yielding three types: collinear (type-I), coplanar (type-II), and noncoplanar (type-III). Symmetries flipping momentum (\(g_i\bm{k}=-\bm{k}\)) determine the parity of spin splitting, yielding odd-parity and even-parity classes. For even-parity NSS, additional operations satisfying \(g_i\bm{k}\neq\pm\bm{k}\) is required to enforce zero magnetization. The resulting spin textures form representations of emergent Laue group \(\tilde{G}\) with odd-parity or even-parity.}
\label{Figs1}
\end{figure*}

In this Letter, we present a unified SSG-based framework for establishing symmetry criteria for both OPMs and EPMs, as illustrated in Fig.~\ref{Figs1}. First, the collection of momentum-preserving symmetries enables a classification of NSS into three types: collinear (type-I), coplanar (type-II), and noncoplanar (type-III)~\cite{XiaoZhenyu2024}. By further incorporating momentum-flipping symmetries, we identify eight distinct symmetry-driven mechanisms that support OPMs [Table~\ref{tab2}]. Applying a similar analysis to the even-parity class, we identify seven distinct symmetry mechanisms for EPMs  [Table~\ref{tabe3}].
Our unified symmetry framework naturally includes some paradigms of unconventional magnetism. For instance,  altermagnets emerge as one specific mechanism of EPMs. Importantly, both OPMs and EPMs can be realized across collinear, coplanar, and noncoplanar magnetic orders, thereby significantly expanding the landscape of unconventional magnetism beyond previously established paradigms. For both OPMs and EPMs, the spin textures form representations of the emergent Laue group derived from the underlying SSG with odd-parity ($p$ or $f$-wave) and even-parity ($d$- or $g$-wave), respectively, which provides a unified method to diagnose the spin-splitting character directly from symmetry considerations. Based on these established symmetry criteria, we identify numerous candidate materials from the Magndata database, which realize some new symmetry-driven mechanism for both EPMs and OPMs. Our work provides a foundational symmetry framework for unconventional magnetism.

\begin{table*}
\centering
\setlength\tabcolsep{6pt}
\renewcommand{\arraystretch}{2}
\caption{Classification of OPMs and symmetry criteria for their emergence. Columns 3, 4, and 5 list the spin texture, symmetry criteria, and  magnetic orders for OPMs, respectively. In collinear and coplanar magnetic orders, symmetries $g_1=\{U_z(\theta)||I\}$ and $g_5=\{TU_{z}(\pi)||I\}$ are typically preserved, respectively. }
\begin{tabular}{|c|c|c|c|c|}
\hline
\multirow{8}{*}{\makecell{odd-\\ \\ parity\\ \\ magnets}}&types & spin textures & symmetry criteria & magnetic orders  \\
\cline{2-5}
 ~&\multirow{5}{*}{type-I} & \multirow{6}{*}{\makecell{$s_z(\bm{k}) = -s_z(-\bm{k})$, $s_{x,y}(\bm{k}) = 0$}} & (i$_{\text{o}}$) $g_1=\{U_z(\theta) ||I | \tau_1\}$  &  coplanar\\
 \cline{4-5}
 ~&~&~& (ii$_{\text{o}}$) $g_1=\{U_z(\theta) ||I | \tau_1\}$ and $g_5=\{TU_{z}(\pi)||\tau_5\}$  &  noncoplanar\\
\cline{4-5}
~&~&~&  (iii$_{\text{o}}$) $g_3=\{T ||I | \tau_3\}$ and $\nexists{\{C_2T||I\}}$   &  collinear  \\
 \cline{4-5}
~&~&~& (iv$_{\text{o}}$) $g_1=\{U_z(\theta) ||I | \tau_1\}$ and $g_3=\{T ||I | \tau_3\}$   & noncoplanar \\
 \cline{4-5}
~&~&~&  (v$_{\text{o}}$) $g_6=\{C_2 ||P \}$ and $\nexists{\{C_2T||I\}}$   &  collinear   \\
 \cline{4-5}
~&~&~& (vi$_{\text{o}}$)  $g_1=\{U_z(\theta) ||I | \tau_1\}$ and $g_6=\{U_{x}(\pi)||P\}$    & \makecell{ 
 noncoplanar} \\
\cline{2-5}
~&type-II &$s_{x,y}(\bm k)=-s_{x,y}(-\bm k)$, $s_{z}(\bm k)=0$ & (vii$_{\text{o}}$) $g_2=\{TU_z(\pi)||P\}$ and $g_3=\{T||I|\tau_3\} $ &  noncoplanar \\
\cline{2-5}
 ~&type-III &$s_{x,y,z}(\bm{k}) = -s_{x,y,z}(-\bm{k})$& (viii$_{\text{o}}$) $g_3=\{T||I|\tau_3\}$ & noncoplanar \\
\hline
\end{tabular}
\label{tab2}
\end{table*}

\textit{Symmetry analysis}.---We begin by examining the constraints of spin group symmetry  on the momentum-space spin texture, defined as \(\bm{S}(\bm{k}) = (s_x(\bm{k}), s_y(\bm{k}), s_z(\bm{k}))\), where \(s_j(\bm{k}) = \langle \psi_{\bm{k}} | \sigma_j | \psi_{\bm{k}} \rangle\) for \(j = x, y, z\). Here, \(\sigma_{x,y,z}\) represent the Pauli matrices in the spin space and \(|\psi_{\bm{k}}\rangle\) denotes the Bloch state at momentum \(\bm{k}\).
 To analyze these constraints, we introduce a spin-space group operation \(g_i = \{X_i U_i || R_i | \tau_i\}\), where \(\{R_i | \tau_i\}\) acts on the spatial coordinates (with \(R_i\) being a point group element and $\tau_i$ a translation), and \(\{X_i U_i\}\) operates on the spin degrees of freedom. Here, \(X_i\) is either the identity operator \(I\) or the time-reversal operator \(T\), \(U_i \in \mathrm{SU}(2)\) is a spin rotation that can be represented by a \(\mathrm{SO}(3)\) operation. Under the action of \(g_i\), \(\bm{k}\) transforms as \(\bm{k} \to \lambda_i R_i \bm{k}\), and the spin operator \(\bm{\sigma}\) transforms according to \(\lambda_i {U}_i\), where \(\lambda_i = +1\) (\(-1\)) corresponds to \(X_i = I\) (\(T\)).  Consequently, the vector \(\bm{S}(\bm{k})\) satisfies the symmetry constraint \cite{JiangYi2024,XiaoZhenyu2024}
\begin{equation}
\bm{S}(\lambda_i R_i \bm{k}) = \lambda_i {U}_i \bm{S}(\bm{k}).
\label{eq1}
\end{equation}
Since translation $\tau_i$ does not contribute to the constraints on \(\bm{S}(\bm{k})\), we take $\tau_i=0$ unless otherwise stated.

For a given SSG \(G = \bigcup_i g_i\), each group element $g_i$ imposes a symmetry constraint on \(\bm{S}(\bm{k})\) via Eq.~\eqref{eq1}. Together, \(\bm{S}(\bm{k})\) forms a representation of the emergent point group \(\tilde{G} = \{s_{i} R_{i} \mid g_i \in G\}\) \cite{XiaoZhenyu2024}. Consequently, the SSG \(G\) provides a complete description of NSS. However, this description is cumbersome and not convenient for a quick diagnosis of OPMs and EPMs. Notably, the definitions of odd- and even-parity NSS involve only the relation between \(\bm{S}(\bm{k})\) and \(\bm{S}(-\bm{k})\): \(\bm{S}(\bm{k}) = -\bm{S}(-\bm{k})\) for odd parity and \(\bm{S}(\bm{k}) = \bm{S}(-\bm{k})\) for even parity with $\bm S(\bm k)\neq 0$. Therefore, the classification and symmetry criteria for these two classes can be fully determined by a collection of symmetries that preserve momentum (\(g_i\bm{k} = \bm{k}\)) or flip momentum (\(g_i\bm{k} = -\bm{k}\)). For the odd-parity class, the momentum-flipping symmetries automatically enforce compensated magnetization, as they pair states at \(\bm{k}\) and \(-\bm{k}\) with opposite spin. Consequently, the symmetry criteria for odd-parity NSS directly coincide with those for OPMs, and we will use these terms interchangeably in the following discussion. In contrast, for the even-parity class, additional symmetries satisfying \(g_i\bm{k} \neq \pm \bm{k}\) are required to enforce zero magnetization for realizing EPMs [Fig.~\ref{Figs1}]. Thus, from this perspective, the realization of an EPM is subject to stricter symmetry constraints than those of an OPM.

We first focus on the symmetries that preserve momentum. These symmetries form a subset denoted by $G_1 = \{ g_i\in G|\lambda_{i} R_{i} \bm{k} = \bm{k}\}$. If \(G_1\) contains the combined symmetry \(PT\) (where \(P\) is spatial inversion) or multiple spin-rotation-translation symmetries with distinct rotation axes, the spin degeneracy is enforced \cite{XiaoZhenyu2024,CheXiaobing2024}. Otherwise, NSS emerges, and the symmetries in \(G_1\) impose constraints on the dimensionality of the spin texture \(\bm{S}(\bm{k})\). Two types of such constraining symmetries are
\begin{align}
    g_1 &= \{ U_{1} \parallel I \mid \tau_{1} \}, \quad 
    g_2 = \{ T U_{2} \parallel P \}.
    \label{eq3}
\end{align}
For our purpose, it suffices to consider $U_1 = U_z(\theta)$ (an arbitrary rotation about the $z$-axis) and $U_2 = U_z(\pi)$ [see details in the Supplemental Materials (SMs)~\cite{supp}]. Under these choices, the symmetry \(g_1\) and \(g_2\) impose the constraints 
\begin{align}
    g_1 &: \quad \bm{S}(\bm{k}) = (0, 0, s_z(\bm{k})), \nonumber\\
    g_2 &: \quad \bm{S}(\bm{k}) = (s_x(\bm{k}), s_y(\bm{k}), 0). \label{g2}
\end{align}
Thus, symmetry \(g_1\) forces the spin texture to be collinear, while \(g_2\) forces it to be coplanar.  When neither \(g_1\) nor \(g_2\) is present in \(G_1\), the spin texture is generally noncoplanar.  We emphasize that symmetry $g_1$ is typically preserved in collinear magnetic orders (spin along the $z$ direction) with $\tau_1=\bm 0$. Based on the dimensionality of the symmetry-allowed spin texture,  the NSS can be classified into three types:
collinear (type-I), coplanar (type-II), and noncoplanar (type-III), as illustrated in Fig.~\ref{Figs1}.


\begin{table*}
\centering
\setlength\tabcolsep{4pt}
\renewcommand{\arraystretch}{2}
\caption{Classification of EPMs and symmetry criteria for their emergence. Columns 3--5 list the spin texture, symmetry criteria, and magnetic orders, respectively. Unlike OPMs, EPMs must additionally satisfy the symmetry-enforced zero magnetization (SEZM) condition. In collinear magnetic orders, symmetries $g_1=\{U_z(\theta)||I\}$ and $g_5^{\prime}=\{TU_x(\pi)||I\}$ are typically preserved. In coplanar magnetic orders, the symmetry $g_5=\{TU_z(\pi)||I\}$ is typically preserved. }
\begin{tabular}{|c|c|c|c|c|}
\hline
\multirow{7}{*}{\makecell{Even-\\ \\ parity\\ \\ magnets}}&types & spin textures & symmetry criteria & magnetic orders  \\
\cline{2-5}
~&\multirow{4}{*}{type-I} & \multirow{4}{*}{$s_z(\bm{k}) = s_z(-\bm{k})$, $s_{x,y}(\bm{k}) = 0$} & (i$_{\text{e}}$)  SEZM (altermagnets)   &  collinear   \\
\cline{4-5}
~&~&~& (ii$_{\text{e}}$) $g_1=\{U_z(\pi) ||I | \tau_1\}$ and $g_5^{\prime}=\{TU_x(\pi) ||I \}$ and SEZM   & coplanar \\
\cline{4-5}
~&~&~& (iii$_{\text{e}}$) $g_1=\{U_z(\pi) ||I | \tau_1\}$ and $g_5^{\prime}=\{TU_x(\pi) ||\tau_5 \}$ and SEZM   & noncoplanar \\
\cline{4-5}
~&~&~& (iv$_{\text{e}}$)  $g_1=\{U_z(\theta) ||I | \tau_1\}$ and $g_4=\{I||P\}$ and SEZM   & \makecell{ 
 noncoplanar} \\
 \cline{4-5}
~&~&~&  (v$_{\text{e}}$) $g_1=\{U_z(\theta) ||I | \tau_1\}$ and $g_6=\{U_z(\pi)||P\}$ and SEZM   & \makecell{ 
 noncoplanar} \\
\cline{2-5}
~&type-II &$s_{x,y}(\bm k)=s_{x,y}(-\bm k)$, $s_{z}(\bm k)=0$ & (vi$_{\text{e}}$)  $g_4=\{I||P\} $ and SEZM  &  coplanar \\
\cline{2-5}
 ~&type-III &$s_{x,y,z}(\bm{k}) = s_{x,y,z}(-\bm{k})$& (vii$_{\text{e}}$) $g_4=\{I||P\}$ and SEZM & noncoplanar \\
\hline
\end{tabular}
\label{tabe3}
\end{table*}

We now analyze the momentum-flipping symmetries, which form the subset \(G_2 = \{ g_i \in G \mid \lambda_{i} R_{i} \bm{k} = -\bm{k} \}\). These symmetries impose constraints between \(\bm{S}(\bm{k})\) and \(\bm{S}(-\bm{k})\),  and thus can fully determine the criteria for 
odd-parity and even-parity NSS. The set \(G_2\) includes the following four types of symmetry:
\begin{align}
g_3 &= \{ T \parallel I \mid \tau_{3} \},\quad 
g_4 = \{ I \parallel P \},\nonumber\\
g_5 &= \{ T U_{5} \parallel I \},\quad 
g_6 = \{ U_{6} \parallel P \}.
\label{eq2}
\end{align}
The symmetries \(g_3\) and \(g_4\) impose the constraints
\beqn
&&g_3: \quad \bm{S}(\bm{k}) = -(s_x(-\bm{k}), s_y(-\bm{k}), s_z(-\bm{k})), \label{g3} 
\label{g3}\\
&&g_4: \quad \bm{S}(\bm{k}) = (s_x(-\bm{k}), s_y(-\bm{k}), s_z(-\bm{k})). \label{g4} 
\eeqn
Thus, \(g_3\) ($g_4$) directly enforces the odd-parity (even-parity) NSS and is not compatible with even-parity (odd-parity) NSS. However, the constraints from \(g_5\) and \(g_6\) depend on the specific choice of the spin rotation $U_{5}$ and $U_6$, allowing them to accommodate either parity through appropriate choice \cite{supp}.

\textit{Symmetry criteria for OPMs}.---
To derive the symmetry criteria for OPMs, without loss of generality, we take \(U_{5} = U_z(\pi)\) and \(U_{6} = U_x(\pi)\), respectively. Under this choice, the constraints are as follows
\begin{align}
g_5 &: \quad \bm{S}(\bm{k}) = (s_x(-\bm{k}), s_y(-\bm{k}), -s_z(-\bm{k})), \nonumber\\
g_6 &: \quad \bm{S}(\bm{k}) = (s_x(-\bm{k}), -s_y(-\bm{k}), -s_z(-\bm{k})). \label{g6}
\end{align}
By combining the constraints from momentum-preserving symmetries [Eq.~\eqref{g2}] and momentum-flipping symmetries [Eqs.~\eqref{g3} and \eqref{g6}], we classify OPMs into three types based on the dimensionality of spin textures (see details in SMs \cite{supp}):
\begin{enumerate}
    \item \textbf{Type-I (Collinear odd-parity):} The spin texture is constrained to be collinear, e.g., \(\bm{S}(\bm{k}) = (0,0,s_z(\bm{k}))\) with \(s_z(\bm{k}) = -s_z(-\bm{k})\). This can be realized through three distinct combinations of symmetry operations: (a) \(g_1\) and \(g_3\); (b) \(g_1\) and \(g_5\); (c) \(g_1\) and \(g_6\). In these cases, the symmetry \(g_1\) forces the spin to align along the \(z\)-axis, while \(g_3\), \(g_4\), or \(g_5\) enforce odd parity for $s_z(\bm k)$. It is noted that the combination of \(g_3\) and \(g_5\) also realize OPMs, corresponding to the original proposal for \(p\)-wave magnets \cite{BirkHellenes2023}. However, it is actually a special instance of case (b) when \(U_{1}=U_z(\pi)\) \cite{supp}.
    

    \item \textbf{Type-II (Coplanar odd-parity):} The spin texture is constrained to be coplanar, e.g., \(\bm{S}(\bm{k}) = (s_x(\bm{k}), s_y(\bm{k}), 0)\) with \(s_{x,y}(\bm{k}) = -s_{x,y}(-\bm{k})\). This can be realized by the combination of  symmetries \(g_2\) and \(g_3\), labeled as case (d). Here, \(g_2\) restricts the spin to the \(xy\)-plane, while \(g_3\) imposes odd parity for $s_{x,y}(\bm k)$.

    \item \textbf{Type-III (Noncoplanar odd-parity):} The spin texture $\bm S(\bm k)$ satisfies $s_{x,y,z}(\bm k)=-s_{x,y,z}(-\bm k)$ with $s_{x,y,z}(\bm k)\neq 0$. This is realized in systems under the sole presence of symmetry \(g_3\), labeled as  case (e).
\end{enumerate}

We identify distinct symmetry-driven mechanisms for OPMs by examining the realization of symmetry cases (a)--(e) in collinear, coplanar, or noncoplanar magnetic orders (see SMs \cite{supp} for details). From cases (a)--(c), we derive six symmetry mechanisms, labeled as (i$_{\text{o}}$)--(vi$_{\text{o}}$), for type‑I OPMs in collinear, coplanar, and noncoplanar magnetic orders. In contrast, cases (d) and (e) can only be realized in noncoplanar magnetic orders, yielding mechanisms (vii$_{\text{o}}$) and (viii$_{\text{o}}$) for type‑II and type‑III OPMs, respectively.
We note that in collinear magnetic orders (spin along the $z$ axis), the symmetry $\{C_2T  \parallel I\}$ with $C_2=U_{x/y}(\pi)$ must be broken for realizing OPMs since it enforces even-parity NSS.  This can be achieved
by introducing complex electron hopping (see details in SMs \cite{supp}), which has been studied in Floquet-engineered systems \cite{2025li,2025zhu,zhuang2025,2025liu}. The classification and symmetry criteria for OPMs
are summarized in Table~\ref{tab2}.



\textit{Symmetry criteria for EPMs}.---
To derive the symmetry criteria for even-parity NSS, 
we consider an alternative choice of spin rotations, yielding the symmetries $g_5^{\prime}=\{TU_x(\pi)||I\}$ and  $g_6^{\prime}=\{U_z(\pi)||P\}$. The symmetry constraints in Eq.~\eqref{g6} become 
\beqn
&& g_5^{\prime}:\quad \bm S(\bm k)=(-s_x(-\bm k), s_y(-\bm k),s_z(-\bm k)),\nonumber\\
&& g_6^{\prime}:\quad \bm S(\bm k)=(-s_x(-\bm k), -s_y(-\bm k),s_z(-\bm k)).
\label{g61}
\eeqn
By combing the symmetry constraints in Eqs.~\eqref{g2}, \eqref{g4}, \eqref{g6}, and \eqref{g61},
we classify even-parity NSS into three types (see details in SMs \cite{supp}):
\begin{enumerate}
    \item \textbf{Type-I (Collinear even-parity):} The spin texture is constrained to be collinear, e.g., $\bm{S}(\bm{k}) = (0,0,s_z(\bm{k}))$ with $s_z(\bm{k}) = s_z(-\bm{k})$. This can be realized in three distinct combinations of symmetries:
    (a$^{\prime}$) $g_1$ and $g_4$;
    (b$^{\prime}$) $g_1$ and $g_5^{\prime}$; (c$^{\prime}$) $g_1$ and $g_6^{\prime}$. The symmetry $g_1$ forces the spin to be along the $z$-axis, while $g_{4}$, $g_5^{\prime}$, or $g_6^{\prime}$ enforces even parity for $s_z(\bm{k})$.
    
    \item \textbf{Type-II (Coplanar even-parity):} The spin texture is constrained to be coplanar, e.g., $\bm{S}(\bm{k}) = (s_x(\bm{k}), s_y(\bm{k}), 0)$ with $s_{x,y}(\bm{k}) = s_{x,y}(-\bm{k})$. This can be realized by the combination of symmetries $g_4$ and $g_5$, labeled as case (d$^{\prime}$).
    
    \item \textbf{Type-III (Noncoplanar even-parity):} The spin texture $\bm S(\bm k)$ satisfies $s_{x,y,z}(\bm k)=s_{x,y,z}(-\bm k)$ with $s_{x,y,z}(\bm k)\neq 0$. This can be realized under the symmetry $g_5$ alone, labeled as case (e$^{\prime}$)
\end{enumerate}
We emphasize that the above symmetry requirements only ensure \(\bm{S}(\bm{k}) = \bm{S}(-\bm{k})\)  and do not yet guarantee the symmetry-enforced zero magnetization,
a defining characteristic of EPMs. This condition is governed by the spin point group \(\mathcal{P}_s\), formed by the set of all spin rotations \(\{X_i U_i\}\) that appear in $G$ \cite{Yuntian2025}. If \(\mathcal{P}_s\) is polar, it allows a net magnetic moment. If \(\mathcal{P}_s\) is nonpolar, it enforces compensated magnetization. Therefore, the realization of an EPM requires the simultaneous satisfaction of two conditions: (i) the symmetry criteria for even-parity NSS derived above, and (ii) a nonpolar spin point group \(\mathcal{P}_s\). 


We enumerate the distinct symmetry-driven mechanisms for EPMs by identifying the realization of symmetry cases (a$^{\prime}$)-(e$^{\prime}$) in different magnetic orders.
 From cases (a\('\))–(c\('\)), we derive five distinct symmetry mechanisms (i\(_{\text{e}}\))–(v\(_{\text{e}}\)) for type‑I EPMs across collinear, coplanar, and noncoplanar magnetic orders. Cases (d\('\)) and (e\('\)), by contrast, can only be realized in noncoplanar magnetic orders, yielding mechanisms (vi\(_{\text{e}}\)) and (vii\(_{\text{e}}\)) for type‑II and type‑III EPMs, respectively. The classification and symmetry criteria of EPMs are summarized in Table~\ref{tabe3}.

We relate our classification to some well-studied systems. In collinear magnetic orders, the symmetries \(g_1\) ($\tau_1=\bm 0$) and \(g_5'\) are typically preserved, enforcing even-parity NSS. When the zero magnetization condition is additionally satisfied (spin-contrasted sublattices are related by rotations or mirrors), a type‑I EPM is realized. This scenario corresponds to the widely studied altermagnets~\cite{ifmmode2022a}.  Type‑II EPMs, characterized by coplanar spin textures, are exemplified by MnSe\(_3\)~\cite{chenhua2014}, which corresponds to the mechanism (vi\(_{\text{e}}\)). Type‑III EPMs, featuring noncoplanar even‑parity spin textures, have been experimentally identified in MnTe\(_2\)~\cite{Zhu2024}, corresponding to the mechanism (vii\(_{\text{e}}\)).

\textit{Candidate materials and diagnosis of NSS}.---Based on the established symmetry criteria, we identify 33 candidate OPM materials using the online tool FINDSPINGROUP~\cite{CheXiaobing2024} together with the Bilbao Crystallographic Server~\cite{Gallego:ks5532}. In addition, we find 63 candidate EPM materials (excluding altermagnets) that satisfy the symmetry criteria. All candidates are listed in Appendix~\hyperlink{A}{A}.
Among these candidates, two materials stand out as realizations of fundamentally new symmetry mechanisms that deviate from previously established paradigms. The material CoCrO\(_4\), recently also predicted in Ref.~\onlinecite{2025song}, realizes the mechanism (ii\(_{\text{e}}\)) for type-I EPMs: it exhibits a collinear even-parity spin texture despite having a coplanar magnetic order. This is in stark contrast to altermagnets, which require collinear magnetic orders to host collinear spin textures. Even more striking is the material Sr\(_2\)Fe\(_3\)Se\(_2\)O\(_3\), which realizes the mechanism (iv\(_{\text{o}}\)) for type-I OPMs. This material hosts a collinear odd-parity spin texture in a noncoplanar magnetic order, fundamentally different from the widely studied OPM paradigms \cite{BirkHellenes2023} that require coplanar magnetic orders. These examples demonstrate the power of our symmetry framework in revealing 
new mechanisms for unconventional magnetism.

In addition to the symmetries that preserve or flip momentum (those in $G_1$ and $G_2$), the remaining symmetries associated with \(\lambda_i R_i\bm{k} \neq \pm \bm{k}\), form the subset \(G_3\). The three subsets \(G_1\), \(G_2\), and \(G_3\) partition the SSG \(G\) which provides a complete description of NSS. Each subset plays a distinct role: \(G_1\) determines the dimensionality of the spin textures; \(G_2\) dictates the parity property of \(\bm{S}(\bm{k})\); and \(G_3\) relates spin textures at \(k_1\) and \(k_2\) with \(k_1\neq \pm k_2\), thus determining the specific partial‑wave channels of NSS.
Collectively, the spin textures \(\bm{S}(\bm{k})\) form a representation of the emergent point group \(\tilde{G}\) \cite{XiaoZhenyu2024}. If the subset \(G_2\) is not empty, the point group \(\tilde{G}\) necessarily contains the inversion operation and therefore belongs to a Laue group. For even‑parity (odd‑parity) NSS, \(\bm{S}(\bm{k})\) transforms as an even‑parity (odd‑parity) representation of \(\tilde{G}\). This reveals the mathematical nature of "parity" for OPMs and EPMs:  the emergent inversion in \(\tilde{G}\) does not require the exact symmetry of inversion \(P\) in systems, allowing parity-defined for $\bm S(\bm k)$ even in inversion-breaking crystals.

Once the representation formed by $\bm S(\bm k)$ is identified, the explicit form of NSS can be derived. As an illustration, considering the candidate CsFeCl\(_3\) (type‑I OPM), its SSG is \( P^{2_{100}} 6_3 / ^1m^{2_{100}}m^1c|(3^1_{001}, 3^1_{001}, 1)^m 1\), and the emergent Laue group is \(D_{6h}\). Applying Eq.~\eqref{eq1} shows that \(s_z(\bm{k})\) transforms as the \(B_{2u}\) representation of \(D_{6h}\), with \(y(y^2-3x^2)\) as the lowest‑order basis function. Expanding near the \(\Gamma\) point, we have \(s_z(\bm{k}) \approx k_y(k_y^2-3k_x^2)\), which exhibits an \(f\)-wave pattern. The explicit forms of NSS for all candidate OPM materials are provided in SMs \cite{supp}, which reveals many novel spin textures.

\textit{Discussion and conclusion}.---We emphasize that our framework also reveals the possibility of more exotic spin-splitting behaviors beyond the pure odd-parity or even-parity types. Under certain symmetry constraints, hybrid-parity spin textures can emerge, where some components of \(\bm{S}(\bm{k})\) exhibit odd-parity splitting, while other components exhibit even-parity splitting. A systematic exploration of these hybrid cases, along with their material realizations, will be studied in our future work.

To validate the established criteria, we construct and analyze two theoretical models of OPMs in the Appendix \hyperlink{B}{B}.  Moreover, due to the existence of effective time-reversal symmetry, OPMs can also be topologically nontrivial. In Appendix \hyperlink{C}{C}, we propose and study a topological OPM with helical edge states, realized in a bilayer breathing kagome lattice with noncoplanar magnetic order.

In summary, we establish a unified SSG framework that provides comprehensive symmetry criteria for both EPMs and OPMs. Our work yields a complete classification of NSS into six types and fifteen distinct symmetry-driven mechanisms [Tables~\ref{tab2} and \ref{tabe3}], fundamentally expanding the landscape of unconventional magnetism beyond the altermagnets paradigm.  Our work lays a foundation for future research in unconventional magnetism.



\section{Acknowledgment}
We acknowledge useful discussions with Zhongyi Zhang, Junwei Liu, Xin Liu, Fengcheng Wu, and Xilin Feng.
K.T.L acknowledges the support from the Ministry of Science and Technology, China, and Hong Kong Research Grant Council through Grants No. 2020YFA0309600, No. RFS2021-6S03,
 No. C6025-19G, No. AoE/P-701/20, No. 16310520,
 No. 16307622, and No. 16309223.

\bibliography{reference}

\clearpage

\appendix

\section{\large \textbf{End Matter}}

\renewcommand{\theequation}{A\arabic{equation}}
\setcounter{equation}{0}

\begin{table*}
\centering
\setlength\tabcolsep{3pt}
\renewcommand{\arraystretch}{2}
\caption{Candidate materials for type-I, type-II, and type-III OPMs/EPM. The corresponding symmetry-driven mechanism for candidate materials is also listed.  }
\begin{tabular}{|c|c|c|c|}
\hline
OPMs/EPMs& Types & mechanism & candidate materials\\
\hline
\multirow{5}{*}{\makecell{OPMs}}&\multirow{3}{*}{\makecell{type-I}}&
(iv$_{\text{o}}$) & Sr$_2$Fe$_3$Se$_2$O$_3$ \\
\cline{3-4}
~&~& (v$_{\text{o}}$)& \makecell{
CsFeCl$_3$, ThMn$_2$, EuIn$_2$As$_2$,
CsMnBr$_3$, RbFeCl$_3$,  RbNiCl$_3$,
Ba$_3$CoSb$_2$O$_9$,\\ CsFe(MoO$_4$)$_2$ Er$_2$Pt, DyBe$_{13}$, TbC$_2$, Ca$_2$Cr$_2$O$_5$,   Tm$_5$Pt$_2$In$_4$,  La$_{1/3}$Ca$_{2/3}$MnO$_3$\\ La$_{3/8}$Ca$_{5/8}$MnO$_3$,KFe(PO$_3$F)$_2$,
NiCr$_2$O$_4$, PrMn$_2$O$_5$,  GdMn$_2$O$_5$,
Sr$_2$FeO$_3$Cl,\\  
CoNb$_2$O$_6$, CeNiAsO,
DyMn$_2$O$_5$,BiMn$_2$O$_5$}\\
\cline{2-4}
~&type-II& (vii$_{\text{o}}$) & Ce$_3$InN \\
\cline{2-4}
~&type-III& (viii$_{\text{o}}$) & \makecell{MgV$_2$O$_4$, Mn$_5$Si$_3$,   Ba(TiO)Cu$_4$(PO$_4$)$_4$,\\
Dy$_2$Co$_3$Al$_9$,  
DyFeWO$_6$,
Ho$_2$Cu$_2$O$_5$, BaFe$_2$Se$_3$} \\
\hline
\multirow{5}{*}{\makecell{EPMs}}&\multirow{3}{*}{\makecell{type-I}}&
(i$_{\text{e}}$) &altermagnets \\
\cline{3-4}
~&~& (ii$_{\text{e}}$) & CoCrO$_4$ \\ 
\cline{2-4}
~&type-II& (vii$_{\text{e}}$) & \makecell{$\mathrm{ScMnO}_3$, $\mathrm{Mn}_2\mathrm{O}_3$, Mn$_2$GeO$_4$, Mn$_3$NiN, Mn$_3$GaN,  CoFePO$_5$, Mn$_3$ZnN,\\ Mn$_3$As,  Cu$_2$(OD)$_3$Cl,  Mn$_3$Ge,  RbFe$_2$F$_6$, Mn$_3$Sn,  Sr$_2$CoOsO$_6$,  Er$_2$Sn$_2$O$_7$,\\  Er$_2$Pt$_2$O$_7$,  CdYb$_2$S$_4$, Li$_2$MnTeO$_6$ NdFeO$_3$,  NdCrO$_3$,  ErCrO$_3$,  SmCrO$_3$,\\  CeFeO$_3$,  NdVO$_3$,  YVO$_3$, Gd$_2$Sn$_2$O$_7$,  ErVO$_3$,  Mn$_3$Ir,  Mn$_3$Pt\\  HoCrWO$_6$, BaCoSiO$_4$, Gd$_2$Pt$_2$O$_7$,  Pr$_2$PdGe$_6$,  Ho$_5$Ni$_2$In$_4$, TmVO$_3$,  Li$_2$MnTeO$_6$} \\
\cline{2-4}
~&type-III& (viii$_{\text{e}}$) & \makecell{ Na$_3$Co(CO$_3$)$_2$Cl, CoSO$_4$, Mn$_2$GeO$_4$, NiS$_2$, NH$_4$Fe$_2$F$_6$,  Co$_2$SiO$_4$,  Fe$_2$SiO$_4$,\\  CoSO$_4$, Cd$_2$Os$_2$O$_7$,  MnTe$_2$, Er$_2$Ti$_2$O$_7$,  Rb$_2$Fe$_2$O(AsO$_4$)$_2$,  Dy$_3$Al$_5$O$_{12}$, NiTe$_2$O$_5$,\\ Nd$_2$Sn$_2$O$_7$, Nd$_2$Hf$_2$O$_7$,  Nd$_2$Zr$_2$O$_7$, Dy$_3$Ga$_5$O$_{12}$,  Ho$_3$Al$_5$O$_{12}$,  Tb$_3$Al$_5$O$_{12}$, Ho$_3$Ga$_5$O$_{12}$, \\Tb$_3$Ga$_5$O$_{12}$,  Eu$_2$Ir$_2$O$_7$, Cd$_2$Os$_2$O$_7$, Er$_2$O$_3$,  LaErO$_3$, Nd$_2$Ir$_2$O$_7$} \\
\hline
\end{tabular}
\label{tab3}
\end{table*}

\hypertarget{A}
{{\emph{Appendix A: Candidate materials}}}.--- In this section, we describe the procedure for identifying candidate materials for OPMs and EPMs. In Ref.~\onlinecite{XiaoZhenyu2024}, the authors utilized momentum‑preserving symmetries to explore the materials exhibiting NSS from the Magndata database, classifying them according to the dimensionality of their spin textures. Moreover, \citet{CheXiaobing2024} developed the online tool FindSpinGroup, which identifies all spin space group operations for materials in the Magndata database. Using this tool, we analyzed the candidate materials with NSS identified in Ref.~\onlinecite{XiaoZhenyu2024} to determine the target spin symmetries relevant to OPMs and EPMs. For EPMs, an additional requirement is that the spin point group $\mathcal{P}_s$ must be nonpolar to ensure zero net magnetization. Following this procedure, we identify 33 candidate materials for OPMs. Among these, 25 belong to type-I, one to type-II, and seven to type-III OPMs. Moreover, we find 63 candidate materials (excluding altermagnets) that satisfy the symmetry criteria for EPMs. Of these EPM candidates, one is type-I, 35 are type-II, and 27 are type-III. All these candidates are listed in Table~\ref{tab3} and the corresponding symmetry-driven mechanism is also listed.
We note that our results are not exhaustive and may miss some candidate materials. For instance, 73 type‑I OPMs have been identified in Ref.~\cite{2025song}. Identifying materials for each of the symmetry mechanisms listed in Tables~\ref{tab2} and \ref{tab3} remains an interesting open question.

\begin{figure}
\centering
\includegraphics[width=3.3in]{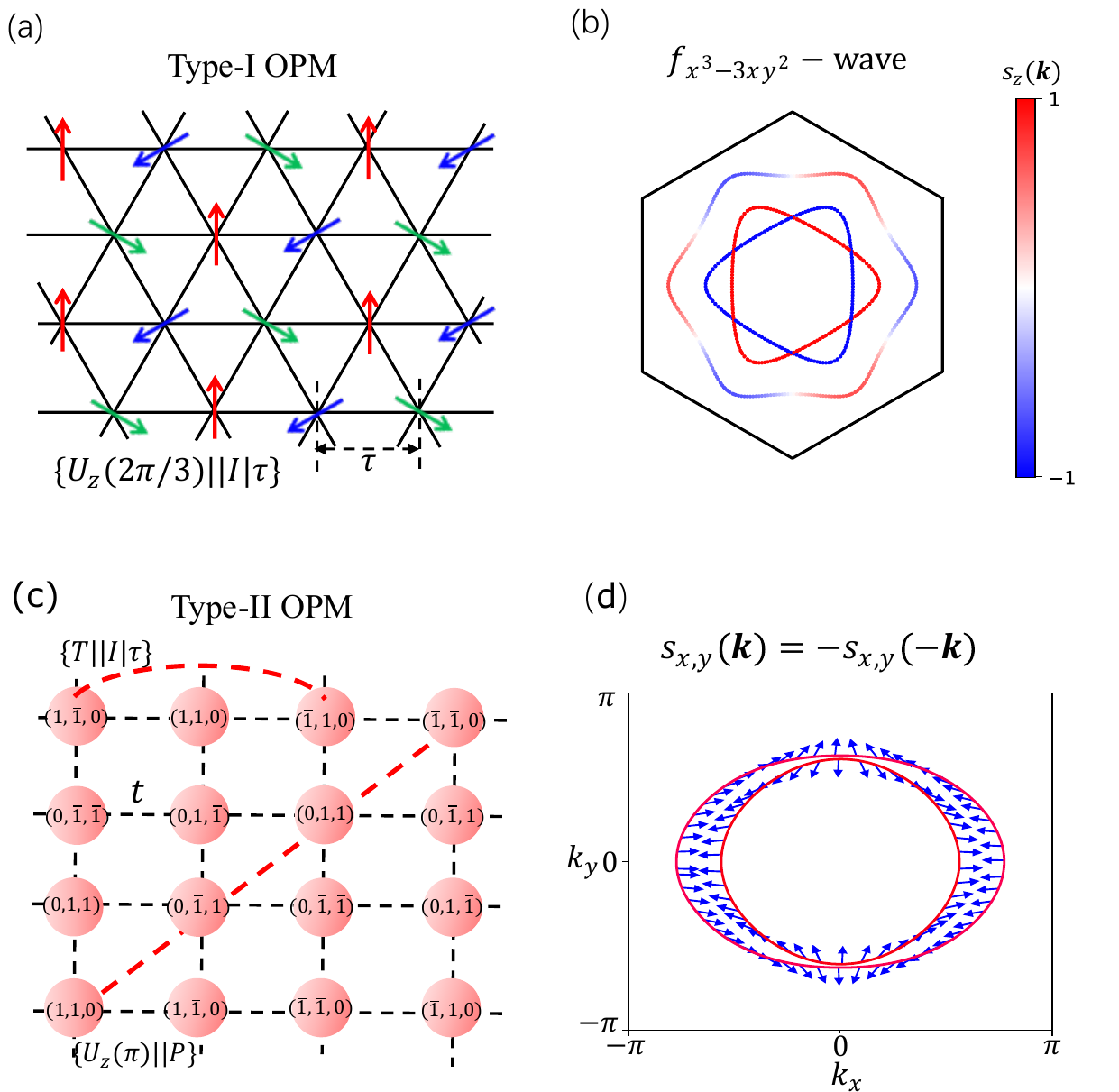}
\caption{ (a) Schematic illustration of $120^{\circ}$ antiferromagnetic order on a triangle lattice. (b) The isoenergy-surface characterized by $s_z(\bm k)$ for the model shown in (a). (c) Magnetic unit cell for a 2D lattice model on a  square lattice. The vectors $(x,y,z)$ denotes the magnetic moment directions. (d) The isoenergy-surface characterized by the vector $(s_x(\bm k,s_y(\bm k))$, shown by the blue arrow, for the lattice model in (c). We take $t=J=1$, $\mu=-2.5$ for (b),  and $\mu=-4.3$ for (d). }
\label{Fig1}
\end{figure}

\hypertarget{B}{{\emph{Appendix B: Theoretical models}}}.---
To validate the symmetry analysis, we construct two theoretical models, $H_1$ and $H_2$, representing type-I and type-II OPMs with coplanar [Fig.~\ref{Fig1}(a)] and noncoplanar [Fig.~\ref{Fig1}(c)] magnetic orders, respectively. The Hamiltonians $H_1$ and $H_2$ are expressed as:
\begin{equation}
H_{1,2} = \sum_{\langle ij\rangle,\sigma} t c_{i\sigma}^\dagger c_{j\sigma} 
+ J \sum_{i,\sigma,\sigma^{\prime}} \bm{m}_i \cdot c_{i\sigma}^\dagger \bm{\sigma}_{\sigma\sigma'} c_{i\sigma'},
\end{equation}
where $c_{i\sigma}^\dagger$ ($c_{i\sigma}$) denotes the creation (annihilation) operator for an electron with spin $\sigma = \uparrow, \downarrow$ at site $i$, and $\langle ij\rangle$ represents the summation over the nearest-neighbor sites with hopping amplitude $t$. The second term describes the exchange coupling  with strength $J$.

The model $H_1$ describes a coplanar $120^\circ$ antiferromagnetic order on a triangular lattice with a $\sqrt{3} \times \sqrt{3}$ structure [Fig.~\ref{Fig1}(a)], comprising three magnetic sublattices (A, B, and C). The corresponding magnetic moments are $\bm{m}_\text{A} = (-\sqrt{3}/2, -1/2, 0)$, $\bm{m}_\text{B} = (\sqrt{3}/2, -1/2, 0)$, and $\bm{m}_\text{C} = (0, 1, 0)$. $H_1$ respects the symmetries $g_1 = \{U_{z}(2\pi/3)|I|\tau\}$ and $g_5 = \{TU_z(\pi)||I\}$, and therefore belongs to the mechanism (i$_{\text{o}}$). Consequently, $H_1$ belongs to type-I OPMs with nonzero $s_z(\bm{k})$. We find that $s_z(\bm{k})$ transforms as a $B_1$ representation of point group $C_{6v}$, with a basis function $x(x^2 - 3y^2)$ \cite{supp}. Consequently, the NSS of $H_1$ can be described by $k_x(k_x^2 - 3k_y^2)\sigma_z$, which is consistent with the numerical verification [Fig.~\ref{Fig1}(b)].

\begin{figure}
\centering
\includegraphics[width=3.3in]{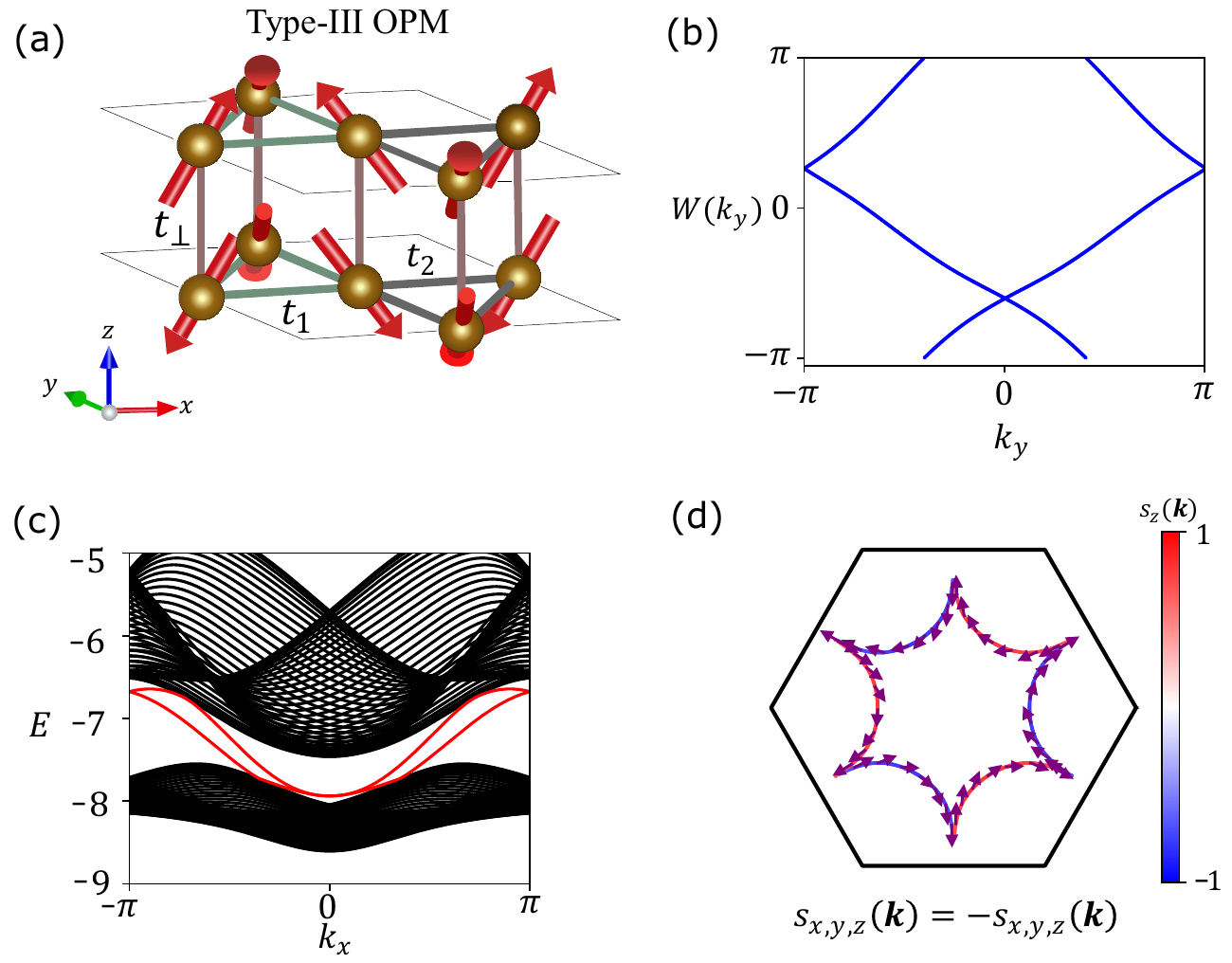}
\caption{(a) Schematic illustration of a bilayer kagome breathing lattice. (b) The wannier center $W(k_y)$ for the lowest energies two bands. (c) The energy spectrum for the system with a nanowire geometry along the $x$ direction. The red bands denote the edge states. (d) The spin polarization Fermi contour. The color encodes the value $s_z(\bm k)$. The pink arrows denote the direction of the vector $(s_x(\bm k), s_y(\bm k))$ at $\bm k$. See SM \cite{supp} for model parameters. }
\label{Fig3}
\end{figure}

The model $H_2$ is defined on a square lattice with sixteen magnetic sublattices. The arrangement of local magnetic moments within a unit cell is illustrated in Fig.~\ref{Fig1}(c). $H_2$ captures the key symmetries of CeIn$_3$N, including $g_3 = \{T||I|\tau\}$ and $g_6 = \{U_z(\pi)||P\}$, and therefore is a type-II OPM with nonzero $s_{x,y}(\bm{k})$, corresponding to the mechanism (vii$_{\text{o}}$). It can be shown that $\{s_x(\bm{k}), s_y(\bm{k})\}$ form a 2D representation $B_1 \oplus B_2$ of the point group $C_{2v}$ \cite{supp}, whose lowest order basis functions are $x\oplus y$. Consequently, the NSS of $H_2$ can be described by $k_x \sigma_x + k_y \sigma_y$ (the superposition coefficients are omitted), which is consistent with the numerical verification [Fig.~\ref{Fig1}(d)]. 

\hypertarget{C}{{\emph{Appendix C: Topological OPMs}}}.---
OPMs can preserve an effective time-reversal symmetry $\tilde{T}$ satisfying $\tilde{T}^2 = -1$, which enables a nontrivial $\mathbb{Z}_2$  classification~\cite{Chiu2016}. To demonstrate such a topological phase, we construct a model on a bilayer breathing Kagome lattice [Fig.~\ref{Fig3}(a)]. This model features layer-contrasted magnetic moments with a finite canting angle, where the in-plane components form an all-in-all-out structure (see  SM \cite{supp} for details). For a single layer, the spin chirality induce Chern band topology and quantum spin Hall states are realized for the bilayer system,  protected by symmetry $ \tilde{T}=\{T || M_z\}$ \cite{XiaoZhenyu2024}. Wilson loop calculations~\cite{YuRui2011} for the two lowest bulk bands confirm their nontrivial $\mathbb{Z}_2$ topology [Fig.~\ref{Fig3}(b)], and Fig.~\ref{Fig3}(c) shows the resulting helical edge states in a nanowire geometry. On the other hand, the inversion symmetry is broken in the breathing Kagome lattice. The whole system behaves as a type-III OPM due to the existence of $\tilde{T}$ symmetry (the same role as $g_3$), corresponding to the mechanism (viii$_{\text{o}}$).  Figure~\ref{Fig3}(d) depicts isoenergy Fermi surfaces characterized by odd-parity spin polarization. Analogously, it can be shown that $\bm S(\bm k)$ form a representation $E_1\bigoplus B_1$ of point group $C_{6v}$ and the NSS can be described by $k_y\sigma_x+k_x\sigma_y+k_x(k_x^2-3k_y^2)\sigma_z$ \cite{supp}.

\newpage
\begin{widetext}
\begin{center}
\begin{large}
\textbf{Supplemental Material for ‘‘Spin Group Symmetry Criteria for Unconventional magnetism"}
\end{large}
\end{center}

\setcounter{figure}{0}
\setcounter{equation}{0}
\renewcommand\thefigure{S\arabic{figure}}
\renewcommand\thetable{S\arabic{table}}
\renewcommand\theequation{S\arabic{equation}}

This Supplemental Material includes the following five sections:
(1) Symmetry criteria for odd-parity magnets (OPMs);
(2) OPMs in collinear magnetic orders;
(3) Symmetry criteria for even-parity magnets (EPMs);
(4) Spin splittings for candidate materials of OPMs.
(5) Theoretical models for OPMs;

\section{A. Symmetry criteria for OPMs}
In this section, we present a detailed derivation of the symmetry criteria for OPMs. These criteria are fully determined by the symmetries that preserve or flip the momentum. We first specify a convenient choice of spin rotations for the key symmetries, which allows us to derive explicit constraints on the spin texture. Under these constraints, we then identify the independent symmetry-driven cases for OPMs. Finally, by examining how these cases are realized in collinear, coplanar, and noncoplanar magnetic orders, we enumerate the distinct symmetry mechanisms for OPMs.

\subsection{A1. Choice of spin rotations in key symmetries}
\label{appendixa1}
As discussed in the main text, five spin group symmetries fully determine the emergence of OPMs. These symmetries are as follows:
\begin{align}
    g_1 &= \{ U_1 \| I | \tau_1\}, \quad 
    g_2 = \{T U_2 \| P \}, \quad 
    g_3 = \{T \| I | \tau_3\}, \quad  
    g_5 = \{T U_5 \| I \},\quad g_6 = \{U_6 \| P \}.
    \label{eq3}
\end{align}
Here, $U_i$ denotes an $\mathrm{SU}(2)$ rotation, which can be represented as a $\mathrm{SO}(3)$ matrix of the form $U_i = R_z(\alpha) R_y(\beta) R_x(\gamma)$, with the rotation matrices given by:
\begin{align}
R_z(\alpha) &= 
 \begin{bmatrix}
\cos \alpha & -\sin \alpha & 0 \\
\sin \alpha & \cos \alpha & 0 \\
0 & 0 & 1
\end{bmatrix}, \quad
R_y(\beta) =
\begin{bmatrix}
\cos \beta & 0 & -\sin \beta \\
0 & 1 & 0 \\
\sin \beta & 0 & \cos \beta
\end{bmatrix}, \quad
R_x(\gamma) =
\begin{bmatrix}
1 & 0 & 0 \\
0 & \cos \gamma & -\sin \gamma \\
0 & \sin \gamma & \cos \gamma
\end{bmatrix},
\label{seq2}
\end{align}
where $\alpha$, $\beta$, and $\gamma$ are rotation angles about the $z$-, $y$-, and $x$-axes, respectively. The spin group symmetries impose constraints on the spin texture $\bm{S}(\bm{k})$ through the relation:
\begin{equation}
\bm{S}(\lambda_i R_i \bm{k}) = \lambda_i {U}_i \bm{S}(\bm{k}),
\label{seq1}
\end{equation}
where $\bm{S}(\bm{k}) = (s_x(\bm{k}), s_y(\bm{k}), s_z(\bm{k}))$ is the spin texture vector, with components $s_j(\bm{k}) = \langle \psi_{\bm{k}} | \sigma_j | \psi_{\bm{k}} \rangle$ for $j = x, y, z$. Here, $\sigma_j$ are Pauli matrices, $|\psi_{\bm{k}}\rangle$ is the Bloch state at momentum $\bm{k}$, and $\lambda_i = \pm 1$ indicates the absence ($+1$) or presence ($-1$) of the time-reversal operation in the symmetry.

The constraint in Eq.~\eqref{seq1} depends on the specific form of $U_i$. For symmetry $g_1$, which leaves $\bm{k}$ invariant, a nonzero rotation around two axes in $U_i$ would enforce $\bm{S}(\bm{k}) = \bm{0}$ according to Eqs.~\eqref{seq2} and \eqref{seq1}. For symmetry $g_2$, one can show that $\bm{S}(\bm{k})$ remains nonzero only when $U_2$ is a $\pi$ rotation about a single axis. Therefore, without loss of generality, we take $U_1 = U_z(\theta)$ (an arbitrary rotation about the $z$-axis) and $U_2 = U_z(\pi)$. For symmetries $g_5$ and $g_6$, which reverse momentum $\bm{k}$, arbitrary rotation angles are formally allowed. However, compatibility with odd-parity  non-relativistic spin splitting (NSS)—which requires at least one component of $\bm{S}(\bm{k})$ to change sign under momentum inversion—restricts $U_6$ to a rotation of $\pi$ and $U_5$ to a rotation about a single axis. Consequently, without loss of generality, we choose $U_5 = U_z(\pi)$ and $U_6 = U_x(\pi)$. With these specific choices, the symmetries become:
\beqn
g_1=\{U_z(\theta)\|I|\tau_1\},\quad  g_2=\{TU_z(\pi)\|I\},\quad g_3=\{T\|I|\tau_3\}, \quad g_5=\{TU_z(\pi)\|I\},\quad g_6=\{U_x(\pi)\|P\},
\eeqn
and they impose the following constraints:
\beqn
&&g_1: \bm S(\bm k)=(0, 0,s_z(\bm k)),\nonumber\\
&& g_2: \bm S(\bm k)=(s_x(\bm k), s_y(\bm k),0),\nonumber\\
&& g_3: \bm S(\bm k)=(-s_x(-\bm k), -s_y(-\bm k),-s_z(-\bm k)),\nonumber\\
&& g_5: \bm S(\bm k)=(s_x(-\bm k), s_y(-\bm k),-s_z(-\bm k))\nonumber\\
&& g_6: \bm S(\bm k)=(s_x(-\bm k), -s_y(-\bm k),-s_z(-\bm k)).
\label{eq4}
\eeqn
From these constraints, we observe that the symmetries \(g_1\) and \(g_2\) force two and one components of \(\bm{S}(\bm{k})\) to vanish, respectively. In contrast, the symmetries \(g_3\), \(g_6\), and \(g_5\) flip three, two, and one components of \(\bm{S}(\bm{k})\), respectively, under \(\bm{k}\to -\bm{k}\). Under these constraints, we classify OPMs into three types based on the dimensionality \(d_s\) of the symmetry-allowed spin texture:
\beqn
&&\text{type-I} \quad(d_s=1): \quad s_{j}(\bm{k}) = -s_j(-\bm{k}), \; s_{m,n}(\bm{k}) = 0;\nonumber\\
&&\text{type-II} \quad (d_s=2): \quad s_{m,n}(\bm{k}) = -s_{m,n}(-\bm{k}), \; s_{j}(\bm{k}) = 0;\nonumber\\
&&\text{type-III} \quad (d_s=3): \quad s_{j,m,n}(\bm{k}) = -s_{j,m,n}(-\bm{k}),
\eeqn
where \(j, m, n \in \{x, y, z\}\) and \(j \neq m \neq n\).

\subsection{A2. Independent symmetry-driven cases for OPMs}

In this subsection, we identify the independent symmetry-driven cases for OPMs by systematically combining the key symmetries derived in A1.
For type-I OPMs (collinear spin texture), the symmetry \(g_1 = \{U_z(\theta)\|I|\tau_1\}\) is essential as it enforces \(s_{x,y}(\bm k)=0\). Combining \(g_1\) with other symmetries yields three independent realizations:
\beqn
&&\text{(a)}\quad g_1 = \{U_z(\theta)\|I|\tau_1\} \text{ and } g_3 = \{T\|I|\tau_3\};\nonumber\\
&&\text{(b)}\quad g_1 = \{U_z(\theta)\|I|\tau_1\} \text{ and } g_5 = \{T U_z(\pi)\|I\};\nonumber\\
&&\text{(c)}\quad g_1 = \{U_z(\theta)\|I|\tau_1\} \text{ and } g_6 = \{U_x(\pi)\|P\}.
\eeqn
Additional combinations can also appear, namely
\beqn
&&\text{(a1)}\quad g_3 = \{T\|I|\tau_3\} \text{ and } g_5 = \{T U_z(\pi)\|I\};\nonumber\\
&&\text{(b1)}\quad g_2^{\prime} = \{T U_x(\pi)\|P\} \text{ (with }U_2 = U_x(\pi)\text{)},\quad g_5 = \{T U_z(\pi)\|I\},\text{ and } g_6 = \{U_x(\pi)\|P\}.
\eeqn
All cases (a)-(c), (a1), and (b1) give rise to \(s_z(\bm k) = -s_z(-\bm k)\) and \(s_{x,y}(\bm k)=0\). However, cases (a1) and (b1) are not independent of cases (a)-(c). For (a1), the product \(g_3 g_5 = \{U_z(\pi)\|I|\tau_3\}\) reproduces \(g_1\), so (a1) reduces to case (b). For (b1), the presence of \(g_2^{\prime}\) and \(g_6\) implies \(T = g_2^{\prime} g_6\). Since time reversal is explicitly broken, a fractional translation must appear in \(g_2\) or \(g_6\); without loss of generality, we assign it to \(g_6 = \{U_x(\pi)\|P|\tau_6\}\). Then the product \(g_2^{\prime} g_5 g_6 = \{U_z(\pi)\|I|\tau_6\}\) gives \(g_1\), showing that (b1) also falls under case (b). Thus, only (a), (b), and (c) are independent for type-I OPMs.

For type-II OPMs (coplanar spin texture), the symmetry \(g_2 = \{T U_z(\pi)\|P\}\) is essential as it enforces \(s_z(\bm k)=0\). Several combinations can give type-II behavior:
\beqn
&&\text{(d)}\quad g_2 = \{T U_z(\pi)\|P\} \text{ and } g_3 = \{T\|I|\tau_3\};\nonumber\\
&&\text{(d1)}\quad g_2 = \{T U_z(\pi)\|P\} \text{ and } g_6' = \{U_z(\pi)\|P\} \text{ (with }U_6 = U_z(\pi)\text{)};\nonumber\\
&&\text{(d2)}\quad g_3 = \{T\|I|\tau_3\} \text{ and } g_6' = \{U_z(\pi)\|P\}.
\eeqn
All the above three cases lead to \(s_{x,y}(\bm k) = -s_{x,y}(-\bm k)\) and \(s_z(\bm k)=0\). However, these are not all independent. For (d1), the product \(g_2 g_6' = T\) implies that a fractional translation must be present; taking \(g_6' = \{U_z(\pi)\|P|\tau_6\}\) gives \(g_3 = g_2 g_6' = \{T\|I|\tau_6\}\), so (d1) reduces to case (d). For (d2), the product \(g_3 g_6' = \{T U_z(\pi)\|P|\tau_3\}\) reproduces \(g_2\), showing that (d2) also falls under case (d). Hence, only one independent case, (d) with \(g_2\) together with \(g_3\), captures all type-II realizations.

For type-III OPMs (noncoplanar spin texture), symmetries \(g_1\) and \(g_2\) must be absent because they would restrict the dimensionality. Type-III OPMs can be realized with the symmetry \(g_3 = \{T\|I|\tau_3\}\) alone; we label this as case (e).
In summary, we have three independent symmetry cases (a)-(c) for type-I OPMs, one independent case (d) for type-II OPMs, and one independent case (e) for type-III OPMs.

\subsection{A3. Derivation of eight symmetry mechanisms for OPMs}
In this subsection, we enumerate the distinct symmetry-driven mechanisms for OPMs by examining how the independent cases (a)–(e) identified in Appendix A2 are realized in collinear, coplanar, and noncoplanar magnetic orders. The  eight resulting mechanisms correspond exactly to those listed in Table~I of the main text.
We first consider the realizations that yield type‑I OPMs, which originate from cases (a), (b), and (c).

\textbf{Case (b):} In coplanar magnetic orders with spins confined to the \(xy\)-plane, the symmetry \(g_5 = \{T U_z(\pi)\|I\}\) is typically preserved. If the symmetry \(g_1 = \{U_z(\theta)\|I|\tau_1\}\) is additionally preserved, a type-I OPM is realized, corresponding to case (b). Cases (a) and (c) also involve \(g_1\), but in coplanar orders the presence of \(g_5\) causes them to reduce to case (b). Consequently, only case (b) gives a distinct mechanism in coplanar orders, corresponding to the
\textbf{mechanism (i\(_{\text{o}}\))} in Table~I.
We note that in collinear magnetic orders along the \(z\) axis,   the symmetry \(g_5=\{TU_z(\pi)||I\}\) must be broken because it flips the spin direction.  In noncoplanar magnetic orders, both \(g_1=\{U_z(\theta)\|I|\tau_1\}\) and \(g_5=\{TU_z(\pi)||I|\tau_5\}\) can also be preserved under certain magnetic structures, yielding the \textbf{mechanism (ii\(_{\text{o}}\))}.

\textbf{Case (a):}  \(g_1 = \{U_z(\theta) || I|\tau_1\}\)  and \(g_3 = \{T \| I | \tau_3\}\).  
This case can be realized in collinear, and noncoplanar magnetic orders, yielding two distinct mechanisms for type-I OPMs.

\begin{itemize}
    \item \textit{Collinear magnetic orders} (spins along the \(z\) axis): Here \(g_1 = \{U_z(\theta) \| I\}\) with \(\tau_1 = 0\) is intrinsically preserved. The symmetry \(g_3 = \{T \| I | \tau_3\}\) implies that the opposite-spin sublattices are related by a fractional translation \(\tau_3\), corresponding to a conventional antiferromagnet. In such collinear magnets, the additional symmetry \(g_5' = \{T U_x(\pi) \| I\}\) is typically preserved. The product \(g_5' g_3 = \{U_x(\pi) \| I | \tau_3\}=g_1^{\prime}\) then generates another spin rotation symmetry \(g_1'\). The pair \(\{g_1, g_1'\}\) enforces spin degeneracy~\cite{CheXiaobing2024}. Therefore, for type-I OPMs to emerge in this case, the symmetry \(g_5'\) must be broken, which can be achieved, for example, by introducing complex electron hopping. We label this mechanism as \textbf{mechanism (iii$_{\text{o}}$)}.

    \item \textit{Noncoplanar magnetic orders}: Both symmetries \(g_1\) and \(g_3\) can be preserved under suitable noncoplanar magnetic structures. We label this mechanism as \textbf{mechanism (iv$_{\text{o}}$)}.
\end{itemize}

\textbf{Case (c):} \(g_1 = \{U_z(\theta) \| I | \tau_1\}\) and \(g_6 = \{U_x(\pi) \| P\}\), type-I OPMs can be realized in collinear and noncopalanr magnetic orders.
\begin{itemize}
    \item \textit{Collinear magnetic orders}: Here \(g_6\) relates opposite-spin sublattices via spatial inversion. In collinear magnets, the symmetry \(g_5' = \{T U_x(\pi) \| I\}\) is typically preserved and the product \(g_5' g_6 = \{T \| P\}\) gives the symmetry \(PT\), which enforces spin degeneracy. Therefore, for type-I OPMs to appear in this case, \(g_5'\) must be broken. We label this mechanism as \textbf{mechanism (v$_{\text{o}}$)}
    \item \textit{Noncoplanar magnetic orders}: Both \(g_1\) and \(g_6\) can be preserved under suitable noncoplanar structures, which produces the \textbf{mechanism (vi)}.
\end{itemize}

\textbf{Case (d):} \(g_2 = \{T U_z(\pi) \| P\}\) and \(g_3 = \{T \| I | \tau_3\}\).  
This combination leads to type-II OPMs with \(s_{x,y}(\bm k) = -s_{x,y}(-\bm k)\). It is incompatible with collinear orders, which can only host collinear spin textures. It is also incompatible with coplanar orders, where the present symmetry $g_3$ instead yields type-I OPMs. Consequently, case (d) can only be realized in noncoplanar magnetic orders, giving rise to \textbf{mechanism (vii)} for type-II OPMs.

\textbf{Case (e):} \(g_3 = \{T \| I | \tau_3\}\) alone.  
This case yields noncoplanar odd-parity spin textures. It is realizable only in noncoplanar magnetic orders; in coplanar orders, \(g_3=\{T||I|\tau_3\}\) symmetry combines with symmetry $g_5=\{TU_z(\pi)||I\}$ give type-I OPMs. We label this mechanism for type-III OPMs as \textbf{mechanism (viii)}.

\begin{figure}
\centering
\includegraphics[width=3.8in]{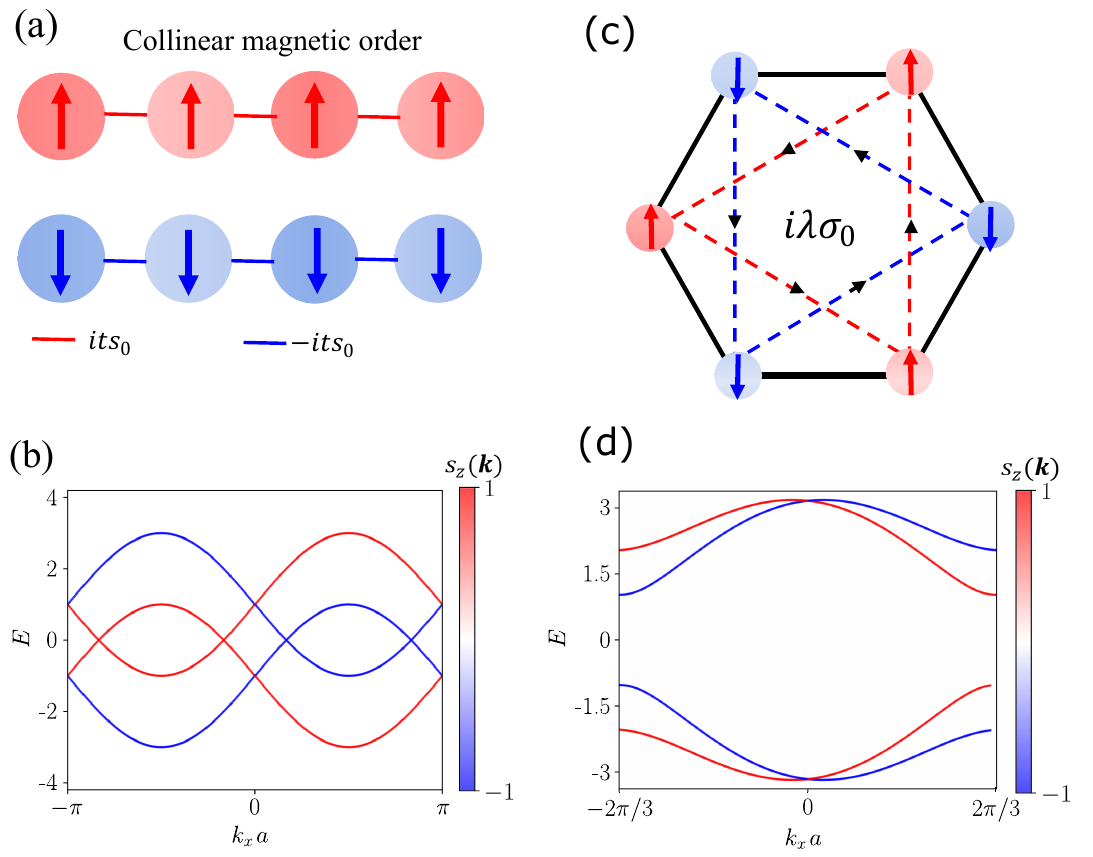}
\caption{ (a) Schematic illustration of a 1D  OPM. (b) The energy bands for the illustrated model in (a). (c) Schematic illustration of Haldane model with antiferromagnet order. (d) The energy bands of the illustrated model in (c). 
 We take $t=J=1$ and $\lambda=0.2$ in (d). }
\label{Fig4}
\end{figure}

\section{C. OPMs in collinear magnetic orders}
\label{AppendixB}
In this section, we demonstrate that OPMs can be realized in systems with collinear magnetic orders and present two theoretical models to illustrate this. In systems with collinear magnetic orders, early studies suggest that only even-parity NSS is allowed, leading to altermagnets~\cite{ifmmode2022, ifmmode2022a}. This is attributed to the effective inversion symmetry $C_2T$ (with $C_2$ being a spin flip operation) for collinear magnetic orders, which flips the momentum \(\bm{k}\) while preserving the spin. However, this argument overlooks the electron hopping, which can break the $C_2T$ symmetry. In the following, we use two theoretical models to demonstrate how OPMs can emerge in such systems.

As illustrated in Fig.~\ref{Fig4}(a), the system consists of two sublattices: sublattice A (red atoms) and sublattice B (blue atoms). We consider the complex hopping for sublattices A and B, which are \(it\) and \(-it\), respectively, and therefore the symmetry \(C_2 T\) is broken. The model Hamiltonian for this one-dimensional system is given by:
\begin{align}
\mathcal{H}_1 = J \tau_z \sigma_z + 2t \sin( k_xa) \tau_z \sigma_0,
\end{align}
where the first term represents the onsite exchange interaction with strength \(J\),  the second term describes the sublattice-dependent complex hopping with magnitude \(t\), and $a$ is a lattice constant. The Pauli matrices \(\tau_z\) act on the orbital degrees of freedom and \(\sigma_0\) is the identity matrix in the spin space. $\mathcal{H}_1$ respects the symmetry $g_6=\{C_2||P\}=\tau_x\sigma_x$, but breaks the symmetry \(C_2 T = \sigma_z K\), corresponding to  mechanism (v$_{\text{o}}$) for type-I OPMs. 
The energy spectrum of $\mathcal{H}_1$ is $E_s=\pm J+2ts\sin k_x$, where $s=1$ ($-1$) for spin up (down). Therefore, the NSS is $E_{+}-E_{-}=4t\sin k_x$, exhibiting a $p_x$-wave pattern. The energy bands of $\mathcal{H}_1$ is plotted in Fig.~\ref{Fig4}(b).

The Haldane model with collinear antiferromagnet order, as illustrated in Fig.~\ref{Fig4}(c), is a typical example for realizing OPMs and was firstly pointed out in Ref.~\onlinecite{YuPing2025}. The model Hamiltonian can be written as 
\beqn
&&\mathcal{H}_2(\bm k)=f_x(\bm k)\tau_x\sigma_0+f_y(\bm{k})\tau_y\sigma_0+f_z(\bm{k})\tau_z\sigma_0+J\tau_z\sigma_z,\nonumber\\
&& f_x(\bm k)=t(1+2\cos (\sqrt{3}k_xa/2)\cos(3k_ya/2)),\nonumber\\
&&f_y(\bm k)=t\cos (\sqrt{3}k_xa/2)\sin(3k_ya/2), \nonumber\\
&&f_z(\bm k)=\lambda(2\sin (\sqrt{3}k_xa)-4\sin (\sqrt{3}k_xa/2)\cos(3k_ya/2)).
\eeqn
where $\lambda$ is the next-nearest-neighbor hopping amplitude. $\mathcal{H}_2$ respects the symmetry $g_3=\{T||I|\tau\}=\tau_x\sigma_yK$ and the symmetry $C_2T$  is broken, corresponding to the mechanism (iii$_{\text{o}}$) for type-I OPMs.  Figure.~\ref{Fig4}(d) plots the energy bands of $\mathcal{H}_2$.

The distribution pattern of $s_z(\bm{k})$ in the momentum space  can be directly revealed from the energy spectrum of $\mathcal{H}_2$, which is 
\beqn
E_{r,s}(\bm k)=r\sqrt{f_x^2(\bm k)+f_y^2(\bm k)+(f_z(\bm k)+sJ)^2},
\eeqn
with $s,r=\pm 1$ with $s=1$ ($s=-1$) for spin up (down).
The energy of the first band of $\mathcal{H}_2$ is 
\beqn
E_1=\sqrt{f_x^2(\bm k)+f_y^2(\bm k)+(|f_z(\bm k)|+J)^2},
\eeqn
which is associated with $s=1$ ($s=-1$) when $f_z(\bm k)>0$ ($f_z(\bm k)<0$). Therefore, $s_z(\bm k)$ for the Bloch states associated with the first energy band has the same sign structure of $f_z(\bm k)$. By expanding $f_z(\bm k)$ around the $\Gamma=(0,0)$ point to the third-order of $\bm k$, we have $f_z(\bm k)\approx -3\sqrt{3}\lambda a^3/4k_x(k_x^2-3k_y^2)$, where we have used the relations $\sin x\approx x-x^3/6$ and $\cos x\approx 1-x^2/2$. Therefore, $s_z(\bm k)\propto  k_x(k_x^2-3k_y^2)$, which exhibits a $f$-wave pattern.

\section{C.  Symmetry criteria for EPMs}
In this section, we present a detailed derivation of the symmetry criteria for EPMs. These criteria are fully determined by the symmetries that preserve or flip momentum. We first specify a convenient choice of spin rotations for the key symmetries, which allows us to derive explicit constraints on the spin texture. Under these constraints, we then identify the independent symmetry-driven cases for even-parity NSS. Finally, by examining how these cases are realized in collinear, coplanar, and noncoplanar magnetic orders, we enumerate the distinct symmetry mechanisms for even-parity NSS, which combines with the condition of symmetry-enforced zero magnetization form the criteria for EPMs.

\subsection{C1. Choice of spin rotations in key symmetries}
As discussed in the main text, five spin group symmetries fully determine the emergence of even-parity NSS. These symmetries are as follows:
\begin{align}
    g_1 &= \{ U_1 \| I | \tau_1\}, \quad 
    g_2 = \{T U_2 \| P \}, \quad 
    g_4 = \{I \| P \}, \quad  
    g_5 = \{T U_5 \| I \},\quad g_6 = \{U_6 \| P \}.
    \label{eqB1}
\end{align}
Following the same spirit, without loss of generality,  we adopt $U_1 = U_z(\theta)$ and $U_2 = U_z(\pi)$. Arbitrary rotations are, in principle, allowed for $U_5$ and $U_6$. However, compatibility with even-parity NSS—which requires at least one component of $\bm{S}(\bm{k})$ to be invariant under $\bm{k}\to -\bm{k}$—restricts $U_5$ to a rotation of $\pi$  and $U_6$ to a rotation about a single axis. Consequently, without loss of generality, we choose $U_5 = U_x(\pi)$ and $U_6 = U_z(\pi)$, and denote the corresponding symmetries as $g_5' = \{T U_x(\pi) \| I\}$ and $g_6' = \{U_z(\pi) \| P\}$. With these specific choices, the symmetries impose the following constraints:
\beqn
&&g_1: \bm S(\bm k)=(0, 0, s_z(\bm k)),\nonumber\\
&& g_2: \bm S(\bm k)=(s_x(\bm k), s_y(\bm k), 0),\nonumber\\
&& g_4: \bm S(\bm k)=(s_x(-\bm k), s_y(-\bm k), s_z(-\bm k)),\nonumber\\
&& g_5': \bm S(\bm k)=(-s_x(-\bm k), s_y(-\bm k), s_z(-\bm k)),\nonumber\\
&& g_6': \bm S(\bm k)=(-s_x(-\bm k), -s_y(-\bm k), s_z(-\bm k)),
\label{eqB4}
\eeqn
Under these constraints, we can classify even-parity NSS into three types based on the dimensionality $d_s$ of the symmetry-allowed spin texture $\bm{S}(\bm{k})$:
\beqn
&&\text{type-I} \quad(d_s=1): \quad s_{j}(\bm{k}) = s_j(-\bm{k}), \; s_{m,n}(\bm{k}) = 0;\nonumber\\
&&\text{type-II} \quad (d_s=2): \quad s_{m,n}(\bm{k}) = s_{m,n}(-\bm{k}), \; s_{j}(\bm{k}) = 0;\nonumber\\
&&\text{type-III} \quad (d_s=3): \quad s_{j,m,n}(\bm{k}) = s_{j,m,n}(-\bm{k}),
\eeqn
where $j$, $m$, and $n$ denote distinct indices in $\{x, y, z\}$.

\subsection{C2. Independent symmetry-driven cases for even-parity NSS}

In this subsection, we identify the independent symmetry-driven cases for even-parity NSS by systematically combining the key symmetries derived in C1.
For type-I even-parity NSS (collinear spin texture),  symmetry \(g_1 = \{U_z(\theta)\|I|\tau_1\}\) is essential, as it enforces \(s_{x,y}(\bm k)=0\). Combining \(g_1\) with other symmetries yields three independent realizations: 
\beqn
&&\text{(a\(^{\prime}\))}\quad g_1 = \{U_z(\theta)\|I|\tau_1\} \text{ and } g_4 = \{I\|P\};\nonumber\\
&&\text{(b\(^{\prime}\))}\quad g_1 = \{U_z(\theta)\|I|\tau_1\} \text{ and } g_5' = \{T U_x(\pi)\|I\};\nonumber\\
&&\text{(c\(^{\prime}\))}\quad g_1 = \{U_z(\theta)\|I|\tau_1\} \text{ and } g_6' = \{U_z(\pi)\|P\}.
\eeqn
Additional combinations can also appear, namely
\beqn
&&\text{(a1\(^{\prime}\))}\quad g_4 = \{I\|P\} \text{ and } g_6^{\prime} = \{U_z(\pi)\|P\};\nonumber\\
&&\text{(b1\(^{\prime}\))}\quad g_2^{\prime} = \{T U_x(\pi)\|P\}, \quad g_5^{\prime} = \{T U_x(\pi)\|I\}, \text{ and } g_6^{\prime} = \{U_z(\pi)\|P\}.
\eeqn
All cases (a\(^{\prime}\))–(c\(^{\prime}\)), (a1\(^{\prime}\)), and (b1\(^{\prime}\)) give rise to \(s_z(\bm k)=s_z(-\bm k)\) and \(s_{x,y}(\bm k)=0\). However, the cases (a1\(^{\prime}\)) and (b1\(^{\prime}\)) are not independent of the cases (a\(^{\prime}\))–(c\(^{\prime}\)). For (a1\(^{\prime}\)), the product \(g_4 g_6^{\prime} = \{U_z(\pi)\|I\}\) reproduces \(g_1\) (with \(\tau_1=\bm 0\)), so (a1\(^{\prime}\)) reduces to case (a\(^{\prime}\)). For (b1\(^{\prime}\)), the product \(g_2^{\prime} g_5^{\prime} g_6^{\prime} = \{U_z(\pi)\|I\}\) again gives \(g_1\), showing that (b1\(^{\prime}\)) also falls under case (a\(^{\prime}\)). Thus, only (a\(^{\prime}\)), (b\(^{\prime}\)), and (c\(^{\prime}\)) are independent for type-I even-parity NSS.

For type-II even-parity NSS (coplanar spin texture), symmetry \(g_2 = \{T U_z(\pi)\|P\}\) is essential, as it enforces \(s_z(\bm k)=0\). Several combinations can give type-II behavior:
\beqn
&&\text{(d\(^{\prime}\))}\quad g_4 = \{I\|P\} \text{ and } g_5 = \{T U_z(\pi)\|I\},\nonumber\\
&&\text{(d1\(^{\prime}\))}\quad g_2 = \{T U_z(\pi)\|P\} \text{ and } g_4 = \{I\|P\};\nonumber\\
&&\text{(d2\(^{\prime}\))}\quad g_2 = \{T U_z(\pi)\|P\} \text{ and } g_5 = \{T U_z(\pi)\|I\}.
\eeqn
All of the above three cases lead to  \(s_{x,y}(\bm k)=s_{x,y}(-\bm k)\) and \(s_{z}(\bm k)=0\). However, these three are equivalent because \(g_2 = g_4 g_5\), \(g_4 = g_2 g_5\), and \(g_5 = g_2 g_4\). Hence, only one independent case, (d\(^{\prime}\)) with \(g_4\) together with \(g_5\), captures all type-II realizations.

For type-III even-parity NSS (noncoplanar spin texture), the symmetries \(g_1\) and \(g_2\) must be absent because they restrict the dimensionality. Type-III even-parity NSS can be realized with the symmetry \(g_4 = \{I\|P\}\) alone; we label this as case (e\(^{\prime}\)).
In summary, we have three independent symmetry cases (a\(^{\prime}\))–(c\(^{\prime}\)) for type-I even-parity NSS, one independent case (d\(^{\prime}\)) for type-II, and one independent case (e\(^{\prime}\)) for type-III.

\subsection{C3. Derivation of seven symmetry mechanisms for even-parity NSS}
In this subsection, we enumerate the distinct symmetry-driven mechanisms for even-parity NSS by examining how the independent cases (a\(^{\prime}\))–(e\(^{\prime}\)) are realized in collinear, coplanar, and noncoplanar magnetic orders. In addition to the parity condition \(\bm S(\bm k)=\bm S(-\bm k)\), the realization of an EPM further requires the symmetry-enforced zero magnetization (SEZM) condition, i.e., the spin point group \(\mathcal{P}_s\) must be nonpolar. The mechanisms listed below already incorporate the necessary symmetries to satisfy the parity condition; the SEZM condition must be checked separately for each candidate material.

We first consider collinear magnetic orders (spins along the \(z\) axis). In such orders, the symmetry \(g_1 = \{U_z(\theta)\|I\}\) with \(\tau_1 = 0\) is intrinsically preserved, and the symmetry \(g_5' = \{T U_x(\pi)\|I\}\) is typically present. These two symmetries together enforce \(s_z(\bm k) = s_z(-\bm k)\) and \(s_{x,y}(\bm k)=0\), i.e., they already guarantee even-parity NSS with a collinear spin texture. Consequently, the only independent case that produces a distinct collinear mechanism is (b\(^{\prime}\)), which gives rise to altermagnets when the SEZM condition is satisfied (spin-contrasted sublattices related by a rotation or mirror). We label this collinear mechanism as \textbf{mechanism (i\(_{\text{e}}\))} in accordance with Table~II. Cases (a\(^{\prime}\)) and (c\(^{\prime}\)) in collinear orders do not introduce additional constraints beyond this intrinsic behavior and are therefore already covered.

We now turn to coplanar magnetic orders. For type-I even-parity NSS, the symmetry \(g_1\) must be present to enforce a collinear spin texture along the \(z\) direction. Coplanar orders with spins confined to the \(xy\)-plane are incompatible with this requirement, as they would instead lead to type-I odd-parity NSS under \(g_1\). The relevant coplanar orders are those where the spin lies in a plane containing the \(z\) axis, such as the \(yz\)-plane. In such orders, the symmetry \(g_5' = \{T U_x(\pi)\|I\}\) is typically preserved. The combination of \(g_1\) and \(g_5'\) then yields type-I even-parity NSS, corresponding to case (b\(^{\prime}\)). Cases (a\(^{\prime}\)) and (c\(^{\prime}\)) also involve \(g_1\) but, in coplanar orders, their effect reduces to that of case (b\(^{\prime}\)) because the presence of \(g_5'\) is generic in these magnetic structures. Consequently, only one distinct mechanism emerges for type-I even-parity NSS in coplanar magnetic orders, which we label as \textbf{mechanism (ii\(_{\text{e}}\))}.

In noncoplanar magnetic orders, there are no spin-only symmetry restrictions such as those in collinear or coplanar orders. Therefore, all three cases (a\(^{\prime}\)), (b\(^{\prime}\)), and (c\(^{\prime}\)) can be realized under suitable magnetic structures, yielding three distinct mechanisms for type-I even-parity NSS:
\begin{itemize}
    \item \textbf{Case (a\(^{\prime}\)) in noncoplanar orders:} \(g_1 = \{U_z(\theta)\|I|\tau_1\}\) and \(g_4 = \{I\|P\}\) enforce a type-I even-parity spin texture along \(z\). This gives \textbf{mechanism (iii\(_{\text{e}}\))}.
    \item \textbf{Case (b\(^{\prime}\)) in noncoplanar orders:} \(g_1 = \{U_z(\theta)\|I|\tau_1\}\) and \(g_5' = \{T U_x(\pi)\|I\}\) produce another type-I even-parity mechanism, labeled \textbf{mechanism (iv\(_{\text{e}}\))}.
    \item \textbf{Case (c\(^{\prime}\)) in noncoplanar orders:} \(g_1 = \{U_z(\theta)\|I|\tau_1\}\) and \(g_6' = \{U_z(\pi)\|P\}\) produce a third type-I even-parity mechanism, labeled \textbf{mechanism (v\(_{\text{e}}\))}.
\end{itemize}

\textbf{Case (d\(^{\prime}\)):} \(g_4 =  \{I\|P\}\) and \(g_5 = \{T U_z(\pi)\|I\}\).  
This combination yields type-II even-parity NSS with \(s_{x,y}(\bm k)=s_{x,y}(-\bm k)\) and \(s_z=0\). Symmetry \(g_5   = \{T U_z(\pi)\|\) is typically preserved in coplanar magnetic orders where the spins lie in the \(xy\)-plane. Therefore, type-II even-parity NSS can be realized in coplanar magnetic orders with inversion symmetry. This corresponds to \textbf{mechanism (vi\(_{\text{e}}\))} for type-II EPMs in Table~II. 
In collinear orders, the spin texture is forced to be collinear, which is incompatible with type-II NSS. Symmetry \(g_5 = \{T U_z(\pi)\|I\}\) is  not compatible with noncoplanar magnetic orders. Thus, case (d\(^{\prime}\))  leads to only one mechanism for EPMs.

 \textbf{Case (e\(^{\prime}\)):} \(g_4 = \{I\|P\}\) alone.  
The presence of \(g_4\) alone imposes \(\bm S(\bm k)=\bm S(-\bm k)\) without restricting the dimensionality, allowing a noncoplanar even-parity spin texture. This case is realizable only in noncoplanar magnetic orders. In collinear magnetic orders, the spin textures must be collinear. In coplanar magnetic orders, the symmetry $g_5=\{TU_z(\pi)||I\}$ is typically preserved, which combines with the symmetry $g_4$ yields type-II even-parity NSS.
Hence, case (e\(^{\prime}\)) is specific to noncoplanar orders and yields \textbf{mechanism (vii\(_{\text{e}}\))} for type-III EPMs.

Collecting the distinct mechanisms, we obtain seven independent symmetry mechanisms for even-parity NSS: (i\(_{\text{e}}\))–(vii\(_{\text{e}}\)), exactly as listed in Table~II.

\section{D. Spin splittings for candidate materials of OPMs}
In this section, we determine the form of spin splitting for the listed candidate OPM materials in Appendix A by symmetry consideration.

For a given material with spin space group \(G\), the corresponding emergent Laue group \(\tilde{G} = \{s_g R_g \mid g \in G\}\) can be derived. The spin texture \(\bm{S}(\bm{k})\) transforms as a \(d_s\)-dimensional representation of \(\tilde{G}\) with odd-parity. The lowest-order basis functions of this representation determine the functional form of \(\bm{S}(\bm{k})\) near the \(\Gamma\) point. For example, the spin space group of the candidate material EuIn\(_2\)As\(_2\) is \(G = P^{6_{001}^1}6_3 / ^{2_{100}}m^1m{^{6_{001}^1}}c(1, 1, 3^1_{001})^m1\). This SSG gives rise to the emergent point group \(\tilde{G} = 6/mmm\). The spin texture component \(s_z(\bm{k})\) transforms as the \(A_{2u}\) representation, whose lowest-order basis function is proportional to \(z\). Therefore, near the \(\Gamma\) point we have \(s_z(\bm{k}) \propto k_z\), and the NSS can be effectively described by the term \(k_z \sigma_z\), which is consistent with the results of the first principle calculation presented in the work \cite{Pari2025}

In Tables~\ref{stab3} and~\ref{stab4}, we summarize the emergent point group \(\tilde{G}\), the representation carried by the spin texture under \(\tilde{G}\), the corresponding basis functions, the functional form of the spin texture near the \(\Gamma\) point, the term \(\bm{S}(\bm{k})\cdot\bm{\sigma}\) describing the NSS, and the associated partial-wave channels for each candidate OPM materials.

\begin{table*}
\centering
\setlength\tabcolsep{5pt}
\renewcommand{\arraystretch}{2.1}
\caption{The emergent point group $\tilde{G}$, the formed representation by $\bm S(\bm k)$ of $\tilde{G}$, the basis function of the formed representation, the functional form of spin texture near the $\Gamma$ point, the term $\bm {S}(\bm k)\cdot \bm{\sigma}$, and the partial wave channel of NSS for each candidate materials. }
\begin{tabular}{|c|c|c|c|c|c|c|c|c|}
\hline
Types&Materials &PG ($\tilde{G}$)&  Rep of $\bm S(\bm k)$& basis function & $\bm S(\bm k)$ & $\bm S(\bm k)\cdot \bm{\sigma}$ & channel\\
\hline
\multirow{14}{*}{\makecell{type-I\\ OPMs}}& Sr$_2$Fe$_3$Se$_2$O$_3$ & $2/m$ & $B_u$ & $x,z$ & $s_y(\bm k)\approx k_x+k_z$ &$(k_x+k_z)\sigma_y$ & $p$-wave\\
\cline{2-8}
~&CsFeCl$_3$& $6/mmm$ & $B_{2u}$ & $y(y^2-3x^2)$ & $s_z(\bm k)\approx k_y(k_y^2-3k_x^2)$ &$k_y(k_y^2-3k_x^2) \sigma_z$ & $f$-wave\\
\cline{2-8}
~& ThMn$_2$ & $6/mmm$ & $B_{2u}$ & $y(y^2-3x^2)$ & $s_z(\bm k)\approx k_y(k_y^2-3k_x^2)$ & $k_y(k_y^2-3k_x^2)\sigma_z$ & $f$-wave\\
\cline{2-8}
~&EuIn$_2$As$_2$&  $6/mmm$ & $A_{2u}$& $z$&  $s_z(\bm k)\approx k_z$& $k_z\sigma_z$ & $p$-wave\\
\cline{2-8}
~&CsMnBr$_3$ & $6/mmm$ & $B_{2u}$ & $y(y^2-3x^2)$ & $s_z(\bm k)\approx k_y(k_y^2-3k_x^2)$ & $k_y(k_y^2-3k_x^2)\sigma_z$ & $f$-wave\\
\cline{2-8}
~&RbFeCl$_3$ & $6/mmm$ & $B_{2u}$ & $y(y^2-3x^2)$ & $s_z(\bm k)\approx k_y(k_y^2-3k_x^2)$ & $k_y(k_y^2-3k_x^2)\sigma_z$ & $f$-wave \\
\cline{2-8}
~&Ba$_3$CoSb$_2$O$_9$ & $6/mmm$ & $B_{2u}$ & $y(y^2-3x^2)$ & $s_y(\bm k)\approx k_y(k_y^2-3k_x^2)$ & $k_y(k_y^2-3k_x^2)\sigma_y$ & $f$-wave \\
\cline{2-8}
~&CsFe(MoO$_4$)$_2$& $\bar{3}$ &$A_{u}$ & $z$ & $s_z(\bm k)\approx k_z$ & $k_z\sigma_z$ & $p$-wave\\
\cline{2-8}
~& Er$_2$Pt & $mmm$ & $B_{2u}$ & $y$& $s_x(\bm k)\approx k_y$ & $k_y\sigma_x$ & $p$-wave\\
\cline{2-8}
~&DyBe$_{13}$ &4/mmm & $A_{2u}$ & $z$ & $s_z(\bm k)\approx k_z$ & $k_z\sigma_z$ & $p$-wave\\
\cline{2-8}
~&TbC$_{2}$ & $mmm$ & $B_{1u}$ & $z$ & $s_z(\bm k)\approx k_z$ & $k_z\sigma_z$ & $p$-wave \\
\cline{2-8}
~&Ca$_2$Cr$_2$O$_5$ & $mmm$ & $B_{1u}$ & $z$ & $s_z(\bm k)\approx k_z$ & $k_z\sigma_z$ & $p$-wave  \\
\cline{2-8}
~& Tm$_5$Pt$_2$In$_4$& $2/m$ & $B_{u}$  &$x,z$ & $s_y(\bm k)\approx k_x+k_z$  & $(k_x+k_z)\sigma_y$ & $p$-wave\\
\cline{2-8}
~& La$_{1/3}$Ca$_{2/3}$MnO$_3$ & $mmm$ & $B_{2u}$ & $y$ &  $s_x(\bm k)\approx k_y$  & $k_y\sigma_x$ & $p$-wave\\
\cline{2-8}
~& La$_{3/8}$Ca$_{5/8}$MnO$_3$ & $mmm$ & $B_{2u}$ & $y$ & $s_x(\bm k)\approx k_y$  & $k_y\sigma_x$ & $p$-wave\\
\cline{2-8}
~& KFe(PO$_3$F)$_2$ & $\bar{3}$ & $A_u$ & $z$ & $s_z(\bm k)\approx k_z$  & $k_z\sigma_z$ & $p$-wave \\
\cline{2-8}
~& NiCr$_2$O$_4$& $D_{2h}$ & $B_{2u}$ & $y$ & $s_y(\bm k)\approx k_y$  & $k_y\sigma_y$ & $p$-wave\\
\cline{2-8}
~&PrMn$_2$O$_5$ & $mmm$ & $B_{2u}$ & $y$ & $s_x(\bm k)\approx k_y$  & $k_y\sigma_x$ & $p$-wave\\
\cline{2-8}
~&GdMn$_2$O$_5$ & $mmm$ & $B_{2u}$ & $y$ & $s_x(\bm k)\approx k_y$  & $k_y\sigma_x$ & $p$-wave \\
\cline{2-8}
~ &Sr$_2$FeO$_3$Cl & $D_{4h}$ & $B_{1u}$ & $xyz$ & $s_z(\bm k)\approx k_xk_yk_z$  & $k_xk_yk_z\sigma_z$ & $p$-wave \\
\hline
\end{tabular}
\label{stab3}
\end{table*}

\begin{table*}
\centering
\setlength\tabcolsep{3pt}
\renewcommand{\arraystretch}{2.1}
\caption{The emergent point group $\tilde{G}$, the formed representation by $\bm S(\bm k)$ of $\tilde{G}$, the basis function of the formed representation, the functional form of spin texture near the $\Gamma$ point, the term $\bm {S}(\bm k)\cdot \bm{\sigma}$, and the partial wave channel of NSS for each candidate materials. }
\begin{tabular}{|c|c|c|c|c|c|c|c|}
\hline
Types&Materials &PG ($\tilde{G}$)&  Rep of $\bm S(\bm k)$& basis function &  $\bm S(\bm k)$ & $\bm S(\bm k)\cdot \bm{\sigma}$ & channel\\
\hline
\multirow{3}{*}{\makecell{type-I\\ OPMs}}~ &CoNb$_2$O$_6$ & $mmm$ & $B_{1u}$ & $z$  & $s_z(\bm k)\approx k_z$  & $k_z\sigma_z$ & $p$-wave\\
\cline{2-8}
~& CeNiAsO & $2/m$ & $B_u$ & $x,z$ & $s_z(\bm k)\approx k_x+k_z$ &$(k_x+k_z)\sigma_z$ & $p$-wave\\
\cline{2-8}
~& DyMn$_2$O$_5$& $mmm$ & $B_{2u}$ & $y$ &  $s_x(\bm k)\approx k_y$ &$k_y\sigma_x$  & $p$-wave \\
\cline{2-8}
~& CeNiAsO & $2/m$ & $B_u$ & $x,z$ & $s_z(\bm k)\approx k_x+k_z$ & $(k_x+k_z)\sigma_z$ & $p$-wave\\
\cline{2-8}
~& BiMn$_2$O$_5$ & $mmm$ & $B_{2u}$ & $y$ & $s_y(\bm k)\approx k_y$ & $k_y\sigma_x$ & $p$-wave\\
\hline
\makecell{type-II\\ OPMs}& Ce$_3$InN & $4/mmm$ & $E_u$ & $(x,y)$ & $\makecell{s_x(\bm k)\approx k_x\\ s_y(\bm k)\approx k_y}$  & $(k_x\sigma_x+k_y\sigma_y)$  & $p$-wave\\
\hline
\multirow{7}{*}{\makecell{type-III\\ OPMs}}& MgV$_2$O$_4$ &  $mmm$ & $B_{3u} \bigoplus B_{2u}  \bigoplus B_{1u}$ & $(x,y,z)$ & $\makecell{s_x(\bm k)\approx k_x\\ s_y(\bm k)\approx k_y\\ s_z(\bm k)\approx k_z} $ & $\makecell{k_x\sigma_x+\\ k_y\sigma_y+\\ k_z\sigma_z}$ & $p$-wave\\
\cline{2-8}
~& Mn$_5$Si$_3$ & $mmm$ & $B_{1u}\bigoplus B_{1u} \bigoplus B_{1u}$ & $(z,z,z)$ & $s_{x,y,z}(\bm k)\approx k_z $& $\makecell{k_z\sigma_x+\\ k_z\sigma_y+\\ k_z\sigma_z}$ & $p$-wave\\
\cline{2-8}
~& Ba(TiO)Cu$_4$(PO$_4$)$_4$ & $4/mmm$ & $E_u \bigoplus A_{2u}$ & $(x,y)\bigoplus (z) $  & $\makecell{s_x(\bm k)\approx k_x\\ s_y(\bm k)\approx k_y\\ s_z(\bm k)\approx k_z} $ &  $\makecell{k_x\sigma_x+\\ k_x\sigma_y+\\k_z\sigma_z}$ & $p$-wave \\
\cline{2-8}
~& Dy$_2$Co$_3$Al$_9$ & $mmm$ & $B_{2u} \bigoplus B_{3u}  \bigoplus A_{u}$ & $(y,x,xyz)$ & $\makecell{s_x(\bm k)\approx k_y\\ s_y(\bm k)\approx k_x\\ s_z(\bm k)\approx k_xk_yk_z} $& $\makecell{k_y\sigma_x+\\
k_x\sigma_y+\\k_xk_yk_z\sigma_z}$ & $\makecell{s_{x,y}(\bm k), p-\text{wave}\\ s_{z}(\bm k), f-\text{wave}}$\\
\cline{2-8}
~&  Ho$_2$Cu$_2$O$_5$ & $2/m$ & $B_{u} \bigoplus A_{u}  \bigoplus B_{u}$ & $(x,z) \bigoplus y \bigoplus (x,z)$ & $\makecell{s_x(\bm k)\approx k_x+k_z\\ s_y(\bm k)\approx k_y\\ s_z(\bm k)\approx k_x+k_z} $& $\makecell{(k_x+k_z)\sigma_x\\ +k_y\sigma_y+\\(k_x+k_z)\sigma_z}$  & $p$-wave\\
\cline{2-8}
~& BaFe$_3$Se$_3$ & $2/m$ & $A_{u} \bigoplus B_{u}  \bigoplus A_{u}$ & $y \bigoplus (x,z) \bigoplus y$ & $\makecell{s_x(\bm k)\approx k_y\\ s_y(\bm k)\approx k_y+k_z\\ s_z(\bm k)\approx k_y} $&  $\makecell{ k_y\sigma_x+\\(k_x+k_z)\sigma_y\\+k_y\sigma_z}$  & $p$-wave\\
\cline{2-8}
~&TbSbTe & $2/m$ & $B_{u} \bigoplus A_{u}  \bigoplus B_{u}$ & $(x,z) \bigoplus y \bigoplus (x,z)$ &  $\makecell{s_x(\bm k)\approx k_x+k_z\\ s_y(\bm k)\approx k_y\\ s_z(\bm k)\approx k_x+k_z} $ & $\makecell{(k_x+k_z)\sigma_x\\ +k_y\sigma_y+\\(k_x+k_z)\sigma_z}$ & $p$-wave\\
\hline
\end{tabular}
\label{stab4}
\end{table*}

\section{E. Theoretical models for OPMs}
In this section, we present the details for the three theoretical models introduced in the end matter of the main text, which realize type-I, type-II, and type-III OPMs, respectively.

\subsection{E1. Theoretical model for type-I OPMs}
We consider the 120° antiferromagnetic order on a triangular lattice [see Fig.~\ref{Fig4}(e)] with a $\sqrt{3} \times \sqrt{3}$ reconstruction. The unit cell comprises three sublattices, denoted A, B, and C. The local magnetic moments on these sublattices are given by $\bm{m}_\text{A} = (-\sqrt{3}/2, -1/2, 0)$, $\bm{m}_\text{B} = (\sqrt{3}/2, -1/2, 0)$, and $\bm{m}_\text{C} = (0, 1, 0)$.
The model Hamiltonian in real space is expressed as:
\beqn
H_1(\bm r)=\sum\limits_{\bm{r},\sigma,\sigma^{\prime}}J\bm{m}(\bm r)\cdot c_{\sigma}^{\dagger}(\bm r) \bm{\sigma} c_{\sigma^{\prime}}(\bm r)+t\sum_{\langle \bm{r}\bm{r}^{\prime}\rangle,\sigma}c_{\sigma}^{\dagger}(\bm r)c_{\sigma}(\bm r^{\prime}),
\eeqn
where $\bm{m}(\bm{r}) = \left( \cos(\bm{K} \cdot \bm{r} + \phi), \sin(\bm{K} \cdot \bm{r} + \phi), 0 \right)$ represents the spatially varying magnetization, $\bm{\sigma} = (\sigma_x, \sigma_y, \sigma_z)$ is the vector of Pauli matrices, $\bm{K} = \frac{4\pi}{3a}(1/2, \sqrt{3}/2)$ is the propagation-vector of antiferromagnet order, and $\phi = -5\pi/6$ is the phase offset. Here,  $\langle \bm{r} \bm{r}^{\prime} \rangle$ denotes the summation over nearest-neighbor sites.

Since $H_1$ respects the symmetries $g_1=\{U_z(2\pi/3)||I|\tau\}$ and $g_5=\{TU_z(\pi)||I\}$, it belongs to a type-I OPM where only the spin expectation value $s_z(\bm k)$ is nonzero. In addition to $g_1$ and $g_5$, $H_1$ also preserves the symmetries $\hat{C}_{3z}=\{I||R_{3z}\}$, $\hat{M}_y=\{I||M_y\}$, and $\hat{C}_{6z}=g_5\hat{C}_{3z}=\{TU_z(\pi)||R_{3z}\}$. The symmetries $\hat{C}_{6z}$ and $\hat{M}_y$ transform the momentum as $\hat{C}_{6z}\bm k=R_{6z}\bm k$ and $\hat{M}_y\bm k=M_y\bm k$, respectively. These two symmetries serve as generators of the point group $\tilde{G}=C_{6v}$. According to Eq.~\eqref{seq1}, we have  the symmetry constraints
\beqn
s_z(\bm k)=-s_z(\hat{C}_{6z}\bm k), \quad s_z(\bm k)=s_z(\hat{M}_y\bm k).
\eeqn
Therefore, the spin texture $s_z(\bm k)$ forms a $B_1$ representation of $C_{6v}$, the lowest-order basis function of which is $x(x^2-3y^2)$. Consequently, near the $\Gamma=(0,0)$ point, we have $s_z(\bm k)\approx k_x(k_x^2-3k_y^2)$.

\subsection{E2. Theoretical model for type-II OPMs}
In the main text, we construct a theoretical model denoted by $H_2$ to realize type-II OPMs. $H_2$ is defined on a square lattice and comprises sixteen magnetic sublattices per unit cell. The Hamiltonian of $H_2$ is given by
\beqn
H_{2} = \sum_{\langle ij\alpha\beta\rangle,\sigma} t c_{i\alpha\sigma}^\dagger c_{j\beta\sigma} 
+ J \sum_{i,\alpha,\sigma,\sigma^{\prime}} \bm{m}_{\alpha} \cdot c_{i\alpha\sigma}^\dagger \bm{\sigma}_{\sigma\sigma'} c_{i\alpha\sigma'},
\eeqn
where $i$ and $j$ label unit cells, $\alpha=1,\cdots,16$ indexes the sublattices, and $\langle ij\alpha\beta\rangle$ denotes the summation over nearest-neighbor sites. The magnetic moment directions for each sublattice are
\beqn
&&\bm{m}_{1}=(1,1,0),\quad \bm{m}_{2}=(1,-1,0),\quad \bm{m}_{3}=(-1,-1,0),\quad \bm{m}_{4}=(-1,1,0),\nonumber\\
&&\bm{m}_{5}=(0,1,1),\quad \bm{m}_{6}=(0,-1,1), \quad \bm{m}_{7}=(0,-1,-1),\quad \bm{m}_{8}=(0,1,-1),\nonumber\\
&&\bm{m}_{9}=(0,-1,-1),\quad \bm{m}_{10}=(0,1,-1),\quad \bm{m}_{11}=(0,1,1),\quad \bm{m}_{12}=(0,-1,1),\nonumber\\
&&\bm{m}_{13}=(1,-1,0),\quad \bm{m}_{14}=(1,1,0),\quad \bm{m}_{15}=(-1,1,0),\quad \bm{m}_{16}=(-1,-1,0).
\eeqn
The arrangement of these magnetic moments is shown in Figure.~1(c). $H_2$ respects the symmetries $g_3 = \{T||I|\tau\}$ and $g_4 = \{U_z(\pi)||P\}$ and therefore belongs to type-II OPMs with nonzero spin textures $s_{x,y}(\bm{k})$. Additionally, $H_2$ respects the symmetries $\hat{M}_y=\{U_x(\pi)||M_y\}$ and $\hat{M}_x=\{U_y(\pi)||M_x\}$. These spin group symmetries generate the point group $\tilde{G}=C_{2v}$, which is generated by $R_{2z}$ and $M_x$. In particular, $s_x(\bm{k})$ and $s_y(\bm{k})$ form $B_1$ and $B_2$ representations of $C_{2v}$, respectively, which are associated with the symmetry constraints
\beqn
&& s_{x}(k_x,k_y)=-s_{x}(-k_x,k_y),\nonumber\\ 
&& s_{y}(k_x,-k_y)=-s_{y}(k_x,-k_y),\nonumber\\
&& s_{x,y}(k_x,k_y)=s_{x,y}(-k_x,-k_y).
\eeqn
The lowest-order basis functions of the $B_1$ and $B_2$ representations are $x$ and $y$, respectively. Consequently, the NSS of $H_2$ can be described by $k_x \sigma_x + k_y \sigma_y$, where the superposition coefficients are omitted.





\subsection{E3. Theoretical models for type-III OPMs}
 The model Hamiltonian of the type-III OPMs defined on a bilayer breathing kagome lattice in the main text is:
\beqn
H_3&=  \sum\limits_{ \alpha\beta\sigma\ell}(t_1\sum\limits_{\langle ij\rangle}c_{i\alpha\sigma\ell}^{\dagger}c_{j\beta\sigma\ell}+ t_2 \sum\limits_{\bar{\langle ij\rangle}}c_{i\alpha\sigma\ell}^{\dagger}c_{j\beta\sigma\ell})\nonumber\\
&+ \sum\limits_{i\alpha\sigma\ell\ell^{\prime}}t_{\perp}c_{i\alpha\sigma\ell}^{\dagger}c_{i\alpha\sigma\ell^{\prime}}+\sum\limits_{i\alpha\sigma\ell}J\bm{m}_{\alpha\ell}\cdot c_{i\alpha\sigma\ell}^{\dagger} \bm{\sigma} c_{i\alpha\sigma\ell},
\eeqn
where $\alpha, \beta = \text{A, B, C}$ denote the three sublattices, and $\ell, \ell' = \mathfrak{b}, \mathfrak{t}$ represent the bottom and top layers, respectively. 
The first and second terms describe intracell and intercell hopping with amplitudes $t_1$ and $t_2$, respectively. The third term represents interlayer hopping with amplitude $t_\perp$ and the fourth term accounts for noncoplanar magnetic orders described by $\bm{m}_{\alpha\ell}$. We set $t_1\neq t_2$ and $\bm{m}_{\alpha\mathfrak{b}} = -\bm{m}_{\alpha\mathfrak{t}}$. $\bm{m}_{\alpha\ell}$ has a canting angle $\theta=\pi/3$ and in-plane components form an all-in–all-out 
 structure, namely $\bm{m}_{\text{A}\mathfrak{t}} = (1/4, \sqrt{3}/4, \sqrt{3}/2)$, $\bm{m}_{\text{B}\mathfrak{t}} = (-1/4, \sqrt{3}/4, \sqrt{3}/2)$, and $\bm{m}_{\text{C}\mathfrak{t}} = (0, 1/2, \sqrt{3}/2)$. We take the model parameters as $t_{\perp}=t_1=1$ and $t_2=J=4$. In Figure.1~(d), $\mu=-3.3$.
 
$H_3$ respects the symmetry $\tilde{T} = \{T || M_z\}$ and therefore belongs to type-III OPMs with nonzero spin expectation values $s_{x,y,z}(\bm k)$. Additionally, $H_3$ preserves the symmetries $\hat{C}_{2y}=\{U_x(\pi)||R_{2y}\}$ and $\hat{C}_{3z}=\{U_z(2\pi/3)||R_{3z}\}$. These spin group symmetries generate the point group $\tilde{G}=C_{6v}$. According to Eq.~\eqref{seq1}, we obtain the following symmetry constraints:
\beqn
&&s_z(R_{6z} \bm{k})=-s_z(\bm k),\quad s_z(k_x,k_y)=-s_z(-k_x,k_y),\nonumber\\
&&s_x(R_{6z}\bm k)=s_x(\bm k)/2-\sqrt{3}s_y(\bm k)/2,\quad s_y(R_{6z}\bm k)=\sqrt{3}s_x(\bm k)/2+s_y(\bm k)/2,\nonumber\\
&&s_x(M_x\bm k)=s_x(\bm k),\quad s_y(M_y\bm k)=-s_y(\bm k).
\eeqn
Consequently, $s_z(\bm k)$ forms a $B_1$ representation of $C_{6v}$ with the basis function $x(x^2-3y^2)$. Meanwhile, the spin textures $s_x(\bm k)$ and $s_y(\bm k)$ form a two-dimensional $E_1$ representation of $C_{6v}$ with basis functions $(y,x)$. Thus, near the $\Gamma$ point, the NSS of $H_3$ can be described by $k_y\sigma_x+k_x\sigma_y+k_x(k_x^2-3k_y^2)\sigma_z$.

\end{widetext}

\end{document}